\RequirePackage[hyphens]{url}
\documentclass[letterpaper,twocolumn,10pt]{article}
\usepackage{usenix-2020-09}
\usepackage[available, functional, reproduced]{usenixbadges}

\usepackage{macros-private}
\usepackage[most]{tcolorbox} 
\usepackage{enumitem}
\setlist{topsep=2pt, leftmargin=*,parsep=2pt}
\usepackage{subcaption}
\usepackage{xparse}
\usepackage{amsmath}
\usepackage{url}
\usepackage{color}
\usepackage{booktabs}
\usepackage{multirow}
\usepackage{tikz}
\usetikzlibrary{shapes.geometric,arrows,decorations.markings, arrows.meta}
\usetikzlibrary{calc}

\usepackage{nopageno} 

\begin{document}

\title{\Large \bf A limited technical background is sufficient for attack-defense tree acceptability}

\author{
{\rm Nathan Daniel Schiele}\\
Leiden University
\and
{\rm Olga Gadyatskaya}\\
Leiden University
}

\maketitle
\begin{abstract}

Attack-defense trees (ADTs) are a prominent graphical threat modeling method that is highly recommended for analyzing and communicating security-related information. Despite this, existing empirical studies of attack trees have established their acceptability only for users with highly technical (computer science) backgrounds while raising questions about their suitability for threat modeling stakeholders with a limited technical background. Our research addresses this gap by investigating the impact of the users' technical background on ADT acceptability in an empirical study.

Our Method Evaluation Model-based study consisted of $n=102$ participants (53 with a strong computer science background and 49 with a limited computer science background) who were asked to complete a series of ADT-related tasks. By analyzing their responses and comparing the results, we reveal that a very limited technical background is sufficient for ADT acceptability. This finding underscores attack trees' viability as a threat modeling method. \end{abstract}


\section{Introduction}
\label{sec:introduction}

Threat modeling has taken an increasingly prominent role in risk assessment and security-oriented design~\cite{andersonSecurityEngineeringGuide2020}, especially in the area of secure software engineering~\cite{yskout2020threat,howard2003inside,apvrille2005secure}. \emph{Attack-defense trees} (ADTs), a graphical component-based representation of attack scenarios, are a highly recommended model for analyzing attacks as well as communicating attack-related information to others in a succinct manner~\cite{andersonSecurityEngineeringGuide2020,ncsc2023attacktrees}. ADTs have been long considered to be a suitable, versatile, and easy-to-use threat modeling approach~\cite{schneierSecretsLiesDigital2000,shostack2014threat,tarandach2020threat,saini2008threat,shevchenko2018threat,apvrille2005secure,stringhini2021adversarial,ncsc2023attacktrees,reversinglabs2024attacktrees}. However, as threat modeling process and results \emph{need to be accessible to people with different backgrounds}~\cite{dev2023models,brunner2020risk}, in order to be effective as a threat modeling method, ADTs must be acceptable for all stakeholders in the software development process, including, among others, security analysts, software engineers, product owners, and managers~\cite{verreydt2024threat}. For a model to be \emph{acceptable}, stakeholders need to be able to use the model efficiently and effectively, as well as perceive the model to be useful and usable~\cite{moodyMethodEvaluationModel2003}.

Thus far, there have been relatively few studies focusing on ADT acceptability. A few studies have directly compared attack trees with other threat models. For example, Opdahl and Sindre~\cite{opdahlExperimentalComparisonAttack2009} and Karpati et al.~\cite{karpatiComparingAttackTrees2014} found that attack trees allowed for better analysis than misuse cases. Broccia \etal~\cite{broccia_assessing_2024,broccia2025evaluating} have recently demonstrated high comprehensibility and acceptability of ADTs for users with a technical background. Lallie \etal~\cite{lallieEmpiricalEvaluationEffectiveness2017} compared fault trees (the precursor to attack trees) and attack graphs, a temporal state-based threat model~\cite{schieleNovelApproachAttack2021}, in a study with participants of different backgrounds. They found that those with a technical background strongly outperformed those without on both models~\cite{lallieEmpiricalEvaluationEffectiveness2017}.

\revised{As threat modeling is an important part of the secure software development lifecycle~\cite{lipner2023inside,verreydt2024threat} with a strong focus on collaboration~\cite{jawad2024m,verreydt2024threat}, it is crucial to understand whether such a popular and recommended method as ADTs is suitable for all involved stakeholders who might have a very limited technical background.}\ To address this gap in the acceptability of ADTs, re-examine the findings from \cite{lallieEmpiricalEvaluationEffectiveness2017}, and guide our research, we formulated the following research questions following the Method Evaluation Model (MEM) as described by Moody~\cite{moodyMethodEvaluationModel2003}:

\begin{itemize}
\setlength{\itemindent}{1.2em}
    \item[\RQ{1}] Is the actual effectiveness of ADTs affected by technical background?
    \item[\RQ{2}] Are the perceived ease of use or perceived usefulness of ADTs affected by technical background?
    \item[\RQ{3}] Is the intention to use ADTs affected by technical background?
    \item[\RQ{4}] Does technical background impact how ADTs are drawn?
\end{itemize}

Our work aims to establish \emph{whether the extent of the technical background affects the ADT acceptability} by conducting a study with student participants from different fields (53 computer science participants; 49 non-computer science participants with a very limited technical background) who complete the same suite of tasks involving using and creating ADTs. Our study is the first to examine ADTs in this context, and, to the best of our knowledge, our study is the first to examine the creative aspect of using any threat model by having participants create an ADT for a scenario of their own choosing and comparing the resulting set of ADTs.

\textbf{Our main findings are}:
\begin{itemize}
    \item \emph{A limited technical (computer science) background is sufficient for the acceptability of ADTs}: Participants of different backgrounds did not show a significant difference in their usage or perceptions of ADTs.

    \item \emph{A creative component in designing ADTs does not appear to be affected by the background}: All self-drawn ADTs fell within the same general limits (in terms of the number of nodes, depth, refinements, etc.), \revised{represent similar types of scenarios, and are of similar quality}, regardless of the background of their authors.
\end{itemize}

Overall, our results strongly support ADTs as a threat modeling tool that is acceptable for threat modeling stakeholders, including those with a very limited technical (computer science) background. We share our study design, training materials, and the anonymized data from the participants in~\cite{zenodo-dataset} to enable further research in this field. 

The remainder of this paper is structured as follows. \revised{We first present the necessary background information on ADTs and threat modeling and summarize the relevant state-of-practice in threat modeling in Sec.~\ref{sec:background}.}\ We then review the related work on empirical studies in Sec.~\ref{sec:empirical_studies}. Sec.~\ref{sec:methodology} presents the methodology of our study. It is followed by Sec.~\ref{sec:results} presenting the study results and answering our four key research questions. We discuss the results in Sec.~\ref{sec:discussion} and acknowledge the study limitations in Sec.~\ref{sec:limitations}. Sec.~\ref{sec:conclusions} concludes this paper.

\section{Threat Modeling and Attack Trees}\label{sec:background}
The notion of \emph{threat modeling} (TM) refers to a process to identify relevant attacks or threats; it typically takes place in the context of software development or security risk management~\cite{xiong2019threat,tuma2018threat}. In the context of software development, TM can refer to a requirements elicitation or design analysis technique~\cite{shostack2008experiences}. Given the diversity of secure software development guidelines~\cite{kudriavtseva2022secure} and security risk management methods~\cite{gritzalis2018exiting} -- and the wide variety of organizational contexts and systems where threat modeling is applied -- there are also many established TM approaches~\cite{shevchenko2018threat,xiong2019threat,mitre2018tm,granata2024systematic,tatam2021review,tuma2018threat}. These methods differ substantially in their focus and process to follow: e.g., STRIDE helps with discovering pertinent security issues during software development, LINDDUN is designed for privacy threats, TARA and Persona non Grata focus on identifying relevant attacker profiles, while PASTA and OCTAVE cover the whole security risk assessment process~\cite{shevchenko2018threat,mitre2018tm}.

\textbf{Attack trees.}
\emph{Attack trees} (sometimes called \emph{threat trees}) were proposed by Bruce Schneier in 1999, inspired by the fault trees model~\cite{schneierAttackTrees1999}. According to Shevchenko et al.~\cite{shevchenko2018threat}, attack trees are one of the oldest and most widely used threat modeling methods that help capture and dissect possible cyber, cyber-physical, or physical attack scenarios. Attack trees are labeled acyclic graphs (trees) in which every node label is either an attacker's goal or an attack component in service of that goal. Each node can have any number of child nodes with a defined relationship, otherwise referred to as a \emph{refinement}, between those nodes. The \OR\ relationship indicates that one child must be completed for the parent to evaluate as complete. The \AND\ relationship indicates that all children must be completed for the parent to evaluate as complete. We note that each parent node can be refined in only one way (either \AND\ or \OR) or have no children at all: such nodes are called \emph{leaf} nodes, and they represent simple attacker's actions that don't need to be further specified. This simple \AND-\OR\ tree model is very versatile and allows representing complex scenarios succinctly~\cite{schneierAttackTrees1999,mauwFoundationsAttackTrees2006,widel2019beyond}. 

\begin{figure}
\includegraphics[width=0.96\linewidth]{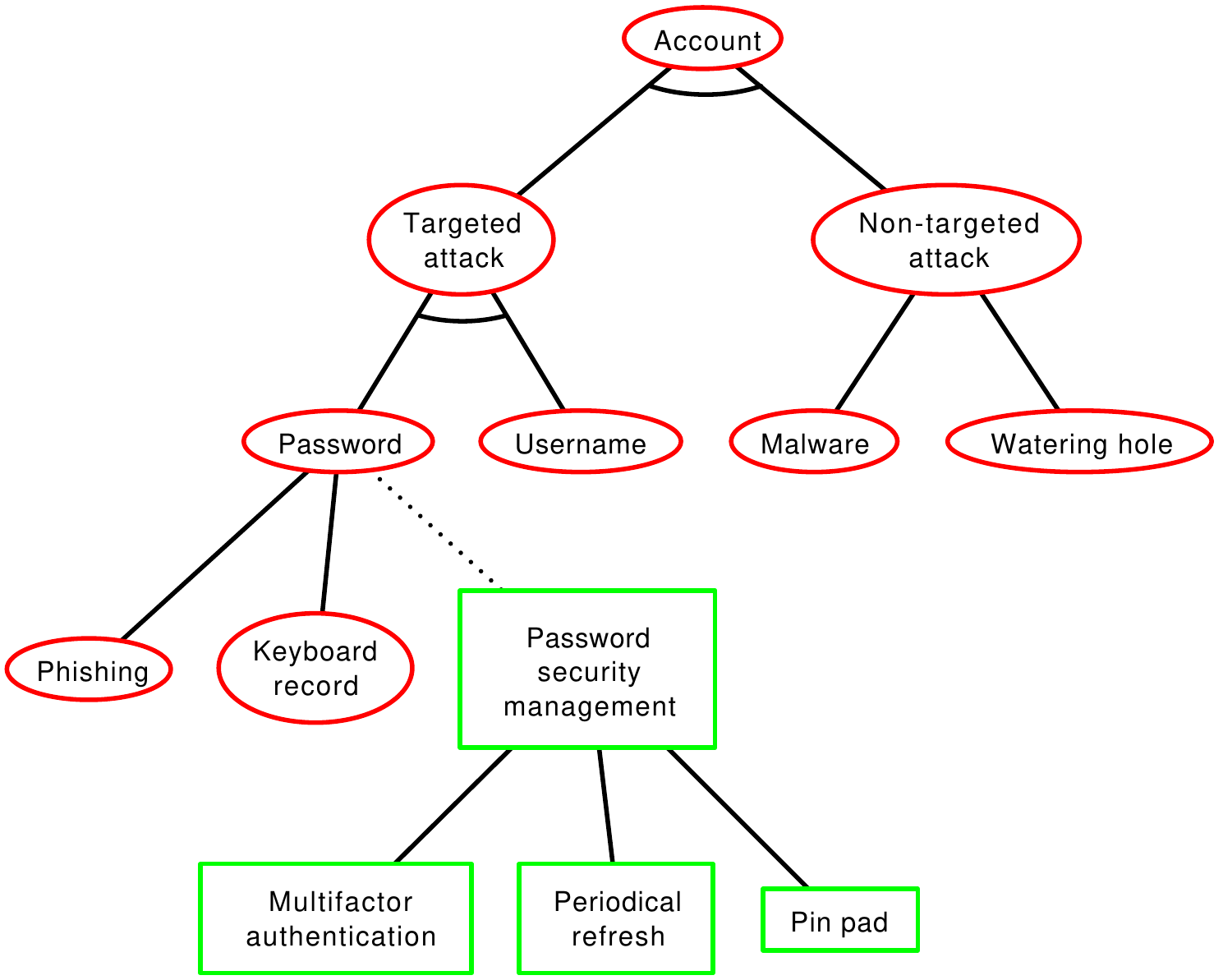}
\caption{An example of an ADT (the second ADT in our small study) from~\cite{sunCyberAttackRisksAnalysis2018}.}
    \label{img:ss-adt2}
\end{figure}

Mauw and Oostdijk defined the attack tree theory by proposing several semantics that can be used to represent attack trees formally~\cite{mauwFoundationsAttackTrees2006}. Kordy~\etal\ further expanded on this by introducing attack-defense trees (ADTs)~\cite{kordyFoundationsAttackDefense2011}. ADTs allow each attack node to have a single countermeasure edge to a defense node, representing a defense to the attack goal or component it is attached to. These defense nodes are roots of their own defense subtree, with the same construction rules as attack trees, including being able to have countermeasure edges to attack nodes, representing an attack against the defense. Thus, attack trees are a particular case of ADTs that do not have any defense nodes. The ADT model allows representing complex attack-defense scenarios where defenders can deploy countermeasures against attacks, and attackers can try to circumvent these countermeasures~\cite{kordyFoundationsAttackDefense2011,kordyDAGBasedAttackDefense2013}. Moreover, considering countermeasures explicitly and  collecting a library of best practices for mitigation are recommended in the TM literature~\cite{dhillon2011developer,jawad2024m,trentinaglia2023eliciting}, and ADTs can help with these objectives.
 Fig.~\ref{img:ss-adt2} shows an example ADT from~\cite{sunCyberAttackRisksAnalysis2018}.

Attack trees and ADTs have been further expanded in other ways~\cite{kordyADToolSecurityAnalysis2013,widel2019beyond,hongSurveyUsabilityPractical2017}. Our work focuses on ADTs 
\emph{{\`a} la} Kordy~\etal~\cite{kordyFoundationsAttackDefense2011}, i.e., without any additional attributes.

\textbf{Attack trees usage in practice.}
Attack trees are quite popular, as evidenced by the fact that they are described in many textbooks (for example, Bishop~\cite{bishop2019computer}, Stallings and Brown~\cite{stallings2018computer}, van Oorschot~\cite{van2020computer}, and Anderson~\cite{andersonSecurityEngineeringGuide2020}), authoritative references on threat modeling (Shostack~\cite{shostack2014threat}, Shevchenko et al.~\cite{shevchenko2018threat}, Bodeau et al.~\cite{mitre2018tm}, and Tarandach and Coles~\cite{tarandach2020threat}), adversarial modeling (CyBOK~\cite{stringhini2021adversarial}), and advice from relevant government institutions and industry bodies (e.g., the UK NCSC~\cite{ncsc2023attacktrees,ncsc2020telecom}), OWASP~\cite{owasp2024tm}, or US NIST~\cite{nist202180030}). 

It is frequently recommended to combine STRIDE-based threat modeling with attack trees for more in-depth analysis of critical data flows and threats~\cite{shostack2014threat,tarandach2020threat,reversinglabs2024attacktrees}. This aligns well with the evidence-based recommendation to complement data flow diagram-based analysis of STRIDE with expressive attacker models by Van Landuyt and Joosen~\cite{van2022descriptivestride} and observations from practitioners that having a library of relevant threat scenarios improves the TM outcomes~\cite{dhillon2011developer}. Schneier advises organizations to develop collections of attack trees to share knowledge and alleviate the need for in-depth security expertise~\cite{schneierAttackTrees1999}. LINDDUN implements this advice, featuring a dedicated privacy threat trees catalogue~\cite{deng2011privacy}, which was appreciated as useful by participants in an empirical study evaluating LINDDUN~\cite{wuyts2014empirical}. Jamil et al.~\cite{jamil2021threat} report that attack trees are chosen as a method because they can help covering all possible attack entry points. 

Despite the popularity, to the best of our knowledge, there are few established references that prescribe how to apply attack trees. Sonderen~\cite{sonderenManualAttackTrees2019} designed a manual for producing attack trees. The manual aims to support a single person designing an attack tree for a given scenario (i.e., the context is not a TM exercise done as a team); it was refined and evaluated in both a qualitative study and a case study. Sonderen reports that careful handling of the levels of abstraction is the most important for a structurally solid attack tree. Schneier prescribes to develop an attack tree top-down, revise it over time, and share with one or more colleagues to improve the completeness of the model~\cite{schneierAttackTrees1999}. He also advises having a library of attack trees that could capture relevant attack scenarios and can be reused -- and thus diminish the need to have security experts around.

\textbf{Threat modeling best practices.}
The TM literature offers substantial insights into the practice of threat modeling. However, it is clear that there is still a gap in understanding how different human factors affect the TM process~\cite{tran2023threat}.
Stevens et al.~\cite{stevens2018battle} reported on their experience with introducing the Center of Gravity TM approach to New York City Cyber Command, highlighting the benefits of threat communication that were reported by the participants. Thompson et al.~\cite{thompson2024there} interviewed and observed 12 medical device security experts to understand their TM practices. They find that the approaches to TM used by different experts vary, and it is important to support a free-flowing, natural approach to ideation (brainstorming).  

Verreydt et al.~\cite{verreydt2024threat} conducted an empirical study of TM methods applied in Dutch organizations within the secure software development process. They found that while the roles involved in the software product development (developers, architects, product owners, and the security team) were central to conducting the TM process itself (this is concurred by other works, e.g.~\cite{shostack2008experiences, bernsmed2022adopting,cruzes2018challenges,dhillon2011developer,trentinaglia2023eliciting}), the outcomes are often communicated to information security officers and managers. Moreover, one of the reasons that management is not involved directly during the TM activities is the belief that such sessions require a strong technical and/or security background~\cite{verreydt2024threat}. Involving a business representative familiar with the key business objectives is also recommended by Ingalsbe et al.~\cite{ingalsbe2008threat}.  Considering security risk management practices, Brunner et al.~\cite{brunner2020risk} also report on the heterogeneity of roles being involved: CxOs, quality and compliance managers, software developers, security-related staff, and others. 

To summarize, TM is a team-based activity that involves different stakeholders: developers, security experts, product owners, and managers. Given the multitude of roles involved, communication becomes very prominent. While TM can be an opportunity to raise awareness about security in managers and bring their attention to the importance of security~\cite{cruzes2018challenges,verreydt2024threat}, difficulties in communication and conveying security messages across the teams are known to be a security ``blocker''~\cite{weir2023incorporating,verreydt2024threat}. Thus, it is important to establish whether such a prominent TM method like attack trees is amenable for all stakeholders, especially for people without a substantial technical background. A positive answer would help organizations to recommend that management and other stakeholders with a limited technical background participate in TM more actively, as well as use attack trees for communicating the TM results outside of the product development team.



\section{Related Work}\label{sec:empirical_studies}

\textbf{Acceptability of attack trees.}
A common strategy for examining TM notations such as ADTs is a study designed to compare two or more notations against each other. Such studies split participants into several groups, and have them complete tasks designed to measure TM method efficacy, with the same tasks being performed using different methods. Opdahl and Sindre used this design to explore the effectiveness of attack trees compared to misuse cases, finding that attack trees are more effective, but the participants has similar perceptions of the two techniques~\cite{opdahlExperimentalComparisonAttack2009}. This study has been replicated with industry practitioners by Karpati~\etal\ who found similar effects, also showing that in the context of cybersecurity, students make a sufficient proxy for practitioners~\cite{karpatiComparingAttackTrees2014}. 

In Diallo~\etal~\cite{dialloComparativeEvaluationThree2006}, two computer science master students applied Common Criteria, misuse cases, and attack trees to the same scenario, evaluating the methods' learnability, usability, analyzability, and clarity of output and finding advantages and disadvantages for each approach. They concluded that attack trees were easy to learn and use, provided a clear output, but were more difficult to analyze~\cite{dialloComparativeEvaluationThree2006}.  

Broccia~\etal\ applied the Method Evaluation Model (MEM) and used 25 human subjects (all with technical background) to examine attack defense tree acceptability~\cite{broccia_assessing_2024}\footnote{This research was performed concurrently with ours, and we had no knowledge of these works when designing and performing our study.}. This study was also recently replicated in another experiment with 49 subjects (computer engineering students)~\cite{broccia2025evaluating}. For their participants with a technical background, Broccia~\etal\ found a good level of understandability and acceptability of ADTs~\cite{broccia_assessing_2024,broccia2025evaluating}. Yet, unlike our study, these studies did not examine participants with a very limited technical background.

To our knowledge, there has only been one previous study on the effect of technical background on attack tree effectiveness. Lallie~\etal\ compared attack graphs to fault trees (considered as a variant of attack trees), finding attack graphs more effective~\cite{lallieEmpiricalEvaluationEffectiveness2017}. Additionally, in the same study, they compared participants with a computer science background to those without one. Their findings did show that computer science participants were able to significantly outperform those without a computer science background using both models. 

\textbf{Studies of other security methods.}
Moving beyond attack trees, Katta~\etal\ conducted an experiment with student participants to compare understanding, performance, and perception of misuse sequence diagrams and misuse case maps, finding that the models perform similarly~\cite{kattaComparingTwoTechniques2010}. 
Labunets~\etal\ compared visual and textual risk assessment methods with student participants using a similar design, focusing on evaluating perception and effectiveness~\cite{labunetsExperimentalComparisonTwo2013,labunets_no_2018}. They found that each type of method was effective in different tasks. De La Vara~\etal\ conducted a study with students concerning Systems Process Engineering Metamodel-like diagrams, comparing this model to text descriptions; they found that the model was statistically significantly more effective in helping students understand the scenario \cite{de_la_vara_empirical_2020}. Tondel et al.~\cite{tondel2019understanding} examined the acceptability of Protection Poker in a study with computer science students and reported that the participants found it to be acceptable but perceived a limited impact on the security of the project. Wuyts et al.~\cite{wuyts2014empirical} empirically evaluated LINDDUN in a series of studies with students and a case study with experts, finding that the method helps to identify relevant privacy threats (correct), but many threats are also not discovered (incomplete). The participants perceived LINDDUN to be easy to use, but the method's efficiency was lower than expected~\cite{wuyts2014empirical}.  

A group of empirical studies focused on evaluating STRIDE-based threat modeling, as STRIDE is the most commonly used method~\cite{verreydt2024threat,thompson2024there,jamil2021threat}. For example, Bernsmed et al.~\cite{bernsmed2022adopting} conducted a study with students to evaluate user acceptance and usage of two versions of a STRIDE-based threat modeling process. Scandariato et al.~\cite{scandariato2015descriptive} evaluated STRIDE in a study with computer science students, concluding that STRIDE is relatively time-consuming (not very efficient), but it is perceived as easy to learn. The threats identified by the participants were largely correct, but many threats were not discovered (low completeness)~\cite{scandariato2015descriptive}. Tuma and Scandariato~\cite{tuma2018two} followed a similar design and compared time cost and effectiveness in threat elicitation of STRIDE per element and STRIDE per interaction in a controlled experiment with computer science master students, reporting that STRIDE per element provided better results. However, all these studies did not examine the effects of participants' background.

\textbf{Examining the difference in backgrounds.}
The existing literature demonstrates that technical background can affect comprehension. Hogganvik and Stolen evaluated the background-affected comprehensibility of risk analysis terminology on professionals and students. They found a statistically significant difference in correct responses, concluding that background does affect comprehension~\cite{hogganvik_risk_2005}. Wu~\etal\ conducted a study examining the ability of participants to understand security texts, finding that a significant percentage of security jargon is not comprehensible by those with a limited IT background~\cite{wu_what_2020}. Chen et al.~\cite{chen2023investigating} found that participants with IT background could understand explanations of Alexa skills privacy policies and related terms better than participants without an IT background.

To summarize, it appears that technical background seems to be an important prerequisite for comprehending many security-related concepts~\cite{lallieEmpiricalEvaluationEffectiveness2017,hogganvik_risk_2005,wu_what_2020}, and, in particular, users without a computer science background might be disadvantaged when using ADTs~\cite{lallieEmpiricalEvaluationEffectiveness2017}. As threat modeling involves participants like managers with possibly a very limited technical background, we set out to examine in our study whether they would be disadvantaged when using ADTs compared to participants with more advanced technical backgrounds.

\section{Methodology}\label{sec:methodology}

This section outlines how we designed the study and collected data to address our research objective.

\subsection{Choosing the methodology}\label{sec:choosing_method}
Examining the aforementioned empirical studies, it is clear that there is no consensus or established guidelines on how to evaluate the acceptability of a threat modeling method such as ADTs, but several key dimensions and methods can be identified. \emph{Comprehensibility} is important when users are presented with some security-relevant information (e.g., privacy policy or a warning)~\cite{wu_what_2020,chen2023investigating}, and it has been a point of attention in empirical studies of security methods~\cite{hogganvik_risk_2005,lallieEmpiricalEvaluationEffectiveness2017}. Moreover, studies examining attack trees and other security risk assessment and threat modeling methods have looked at \emph{effectiveness} in eliciting threats/requirements~\cite{opdahlExperimentalComparisonAttack2009,tuma2018two,labunets2017model}, and \emph{acceptability} for the intended users~\cite{tondel2019understanding,broccia_assessing_2024,broccia2025evaluating}. Note that these objectives are not independent, for example, the comprehensibility of models (how well can users interpret them) produced can be considered as a part of the method's effectiveness~\cite{abrahao2011evaluating,labunets2017model,kattaComparingTwoTechniques2010}, while the effectiveness can be assessed as a component of its acceptability~\cite{broccia_assessing_2024,labunetsFirstEmpiricalEvaluation2014a,stevens2018battle,broccia2025evaluating}.

Two prominent frameworks have been used in the literature to assess the acceptability of a method by its intended users: the Technology Acceptance Model (TAM)~\cite{davisPerceivedUsefulnessPerceived1989} and the Method Evaluation Model (MEM)~\cite{moodyMethodEvaluationModel2003}. TAM focuses on the perceptions of the intended users and it prescribes to measure \emph{perceived usefulness} (\textbf{PU}), \emph{perceived ease of use} (\textbf{PEOU}), and the \emph{intention to use} (\textbf{ITU})~\cite{davisPerceivedUsefulnessPerceived1989}. MEM, depicted schematically Figure~\ref{fig:mem}, extends TAM with components related to actual usage: in addition to the TAM constructs, it recommends measuring \emph{actual effectiveness} (\textbf{AE}), \emph{actual efficiency}  -- and, combined, these constructs will translate into \emph{actual usage}~\cite{moodyMethodEvaluationModel2003}. These two frameworks have been used for evaluating tools and methods in a variety of fields, including cybersecurity. Among the previously mentioned studies, TAM has been applied in, for example, \cite{tondel2019understanding,opdahlExperimentalComparisonAttack2009,karpatiComparingAttackTrees2014,kattaComparingTwoTechniques2010,bernsmed2022adopting}, while MEM was used in~\cite{labunetsFirstEmpiricalEvaluation2014a,labunetsSecurityRiskAssessment,stevens2018battle,broccia_assessing_2024,broccia2025evaluating}. We wish to evaluate the suitability of ADTs as a threat modeling method through the lens of technical background, examining if performance and perceptions change based on the extent of the technical background of users. Since MEM examines perceptions as they relate to actual usage, we believe this framework is a suitable basis for our study design.

Moreover, threat modeling is a creative activity: teams frequently engage in brainstorming~\cite{brunner2020risk} and free-flowing creative thought needs to be facilitated~\cite{thompson2024there}.
Therefore, it is important to examine to what extent threat modeling stakeholders with a very limited technical background might be disadvantaged if they use ADTs for creatively expressing their ideas of relevant attacks. Therefore, to the MEM constructs \textbf{AU}, \textbf{PU} \& \textbf{PEOU}, and \textbf{ITU} (which correspond, respectively, to our \RQ{1}, \RQ{2}, and \RQ{3}) we add another dimension captured by our \RQ{4}. 

We detail how we use the MEM constructs and \RQ{4} in our study context in the remainder of this section.

\subsection{Study design}
\label{ssec:methodology-study-design}

\begin{figure}
\resizebox*{\columnwidth}{!}{\tikzstyle{memnode} = [circle,
minimum size=2.5cm,
text centered,
draw=black,
fill=white]

\tikzstyle{memnodeinvis} = [circle,
minimum size=2.5cm,
text centered,
draw=white,
text=white]

\definecolor{icscolor}{RGB}{31,119,180}
\definecolor{seccolor}{RGB}{255,127,14}
\definecolor{thirdcolor}{RGB}{44,160,44}




\tikzstyle{arrow} = [ultra thick,-latex]

\begin{tikzpicture}[node distance=3cm,]

\node (AE) [memnodeinvis, xshift=-3cm] {\small\shortstack{Actual\\Efficiency}};
\node (AEff) [memnodeinvis, below of=AE] {\small\shortstack{Actual\\Efficacy\\\hypothesis{\hypoCheckUnderstand}, \hypothesis{\hypoSecondADT}, \hypothesis{\hypoErrorAmount}} };
\node (PEOU) [memnodeinvis, right of=AE, xshift=.75cm] {\small\shortstack{Percieved\\Ease of\\Use\\\hypothesis{\hypoSelfUnderstand}, \hypothesis{\hypoWrittenComparison}} };
\node (PU) [memnodeinvis, below of=PEOU] {\small\shortstack{Percieved\\Usefulness\\\hypothesis{\hypoCommunicationTool}, \hypothesis{\hypoAnalysisTool}} };
\node (IU) [memnodeinvis, right of=PEOU, yshift=-1.5cm, xshift=.75cm] {\small\shortstack{Intention\\to Use\\\hypothesis{\hypoIntentionToUse}, \hypothesis{\hypoWrittenComparison}} };
\node (AU) [memnodeinvis, right of=IU, xshift=.75cm ] {\small\shortstack{Actual\\Usage} };

    \Large
\draw[black!50,thick] ($(AE.north west)+(-1,0.6)$)  rectangle  ($(AEff.south east)+(1,-0.6)$) node[anchor=south, pos=0.5, yshift=3cm] {\textbf{Performance}};
\draw[black!50,thick, fill=black!30] ($(PEOU.north west)+(-1,0.6)$)  rectangle ($(PU.south east)+(1,-0.6)$) node[anchor=south, pos=0.5, yshift=2.925cm] {\textbf{Perceptions}};
\draw[black!50,thick] ($(IU.north west)+(-1,2.1)$)  rectangle ($(IU.south east)+(1,-2.1)$) node[anchor=south, pos=0.5, yshift=3cm] {\textbf{Intentions}};
\draw[black!50,thick, fill=black!30] ($(AU.north west)+(-1,2.1)$)  rectangle ($(AU.south east)+(1,-2.1)$) node[anchor=south, pos=0.5, yshift=3cm] {\textbf{Behavior}};

    \draw [arrow] (AE) -- (PEOU);
    \draw [arrow] (AEff) -- (PU);
    \draw [arrow] (PEOU) -- (PU);
    \draw [arrow] (PEOU) -- (IU);
    \draw [arrow] (PU) -- (IU);
    \draw [arrow] (IU) -- (AU);

\node (AE) [memnode, xshift=-3cm] {\small\shortstack{Actual\\Efficiency}};
\node (AEff) [memnode, below of=AE] {\small\shortstack{Actual\\\revised{Effectiveness}\\\hypothesis{\hypoCheckUnderstand}, \hypothesis{\hypoSecondADT}, \hypothesis{\hypoErrorAmount}} };
\node (PEOU) [memnode, right of=AE, xshift=.75cm] {\small\shortstack{Percieved\\Ease of\\Use\\\hypothesis{\hypoSelfUnderstand}, \hypothesis{\hypoWrittenComparison}} };
\node (PU) [memnode, below of=PEOU] {\small\shortstack{Percieved\\Usefulness\\\hypothesis{\hypoCommunicationTool}, \hypothesis{\hypoAnalysisTool}} };
\node (IU) [memnode, right of=PEOU, yshift=-1.5cm, xshift=.75cm] {\small\shortstack{Intention\\to Use\\ \hypothesis{\hypoIntentionToUse}} };
\node (AU) [memnode, right of=IU, xshift=.75cm] {\small\shortstack{Actual\\Usage} };

\end{tikzpicture}}
\caption{The Method Evaluation Model (MEM)~\cite{moodyMethodEvaluationModel2003} with our study hypotheses placed in context.}
\label{fig:mem}
\end{figure}
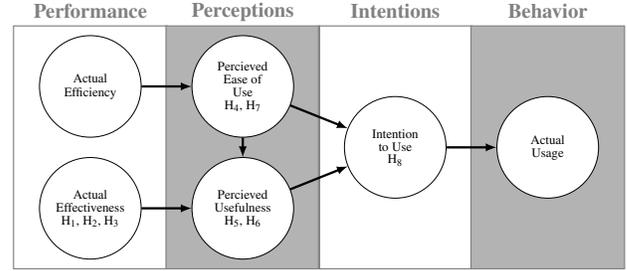

From the research questions \RQ{1}--\RQ{3} derived from the MEM components and our additional \textbf{RQ4} that examines differences in creative usage of ADTs depending on the background, we developed a \textbf{series of hypotheses} to specifically test the aspects of MEM and the creative usage component with the added context of technical background. The hypotheses are presented in Table~\ref{tab:hypotheses}. Figure~\ref{fig:mem} also shows the hypotheses positioned in their relevant MEM component. We have two hypotheses for each aspect we measure: a null hypothesis where we expect no difference between the two groups of participants, and an alternative hypothesis where we expect a difference. We start by testing for a difference as the previous study by Lallie~\etal~\cite{lallieEmpiricalEvaluationEffectiveness2017} observed an influence of technical background on successfully using attack trees.

We measure the actual effectiveness (\textbf{AE}) of ADTs by looking at how well the participants can understand the provided ADT models (\hypothesis{\hypoCheckUnderstand}), how effectively can they design ADTs (\hypothesis{\hypoSecondADT})
, and how many errors they make when designing these ADTs (\hypothesis{\hypoErrorAmount}). We evaluate perceived usefulness (\textbf{PU}) by asking the participants to evaluate on the Likert scale how useful do they find ADTs, separately as a means of threat analysis (\hypothesis{\hypoAnalysisTool}) as well as a means of communication (\hypothesis{\hypoCommunicationTool}), as we want to see whether our participants would demonstrate different preferences depending on the background. We measure perceived ease of use (\textbf{PEOU}) by asking the participants to report on the Likert scale whether they find the provided ADT easy to understand (\hypothesis{\hypoSelfUnderstand}) and if they find it easier to understand a given ADT compared to a textual description (\hypothesis{\hypoWrittenComparison}). Intention to use (\textbf{ITU}) is measured by asking the participants whether they would like to use ADTs in the future, on the Likert scale (\hypothesis{\hypoIntentionToUse}). 

Finally, we studied the creative aspects of designing ADTs by examining how effectively the participants can design ADTs for the self-selected scenario (\hypothesis{\hypoThirdADTsub}) and by measuring, qualitatively and quantitatively, the differences in the ADT models designed for the self-selected scenario (\hypothesis{\hypoThirdADT}).  Specific questions used to measure these aspects are listed in Table~\ref{tab:hypotheses} (text of the referenced questions is available in Appendices~\ref{sec:small-study-q}~and~\ref{sec:large-study-q}.). We provide more details about the measurements done per each hypothesis in the next section (Sec.~\ref{sec:results}). 

Note that we do not measure the actual efficiency of using ADTs separately, because of the study constraints: it was given as a part of a homework assignment where participants worked at their own pace and according to their own schedule. However, we believe we are still able to evaluate ADTs within the scope of the MEM without assessing actual efficiency separately, as the ability of participants to \emph{understand} ADTs by correctly interpreting existing models and creating new ones after a short training translates into both effectiveness and efficiency (see Broccia~\etal~\cite{broccia_assessing_2024}).

\begin{table*}[t!]
\caption{Hypotheses to be investigated by our research and the research questions they contribute to. The null hypothesis in each case proposes no difference, while the alternative hypothesis proposes a difference between the Limited Technical $($\ICS$)$ and Highly Technical $($\SEC$)$ groups. In the measurement questions, \texttt{SS-} refers to questions from the small study, and \texttt{LS-} refers to questions from the large study. Question text can be found in Appendices~\ref{sec:small-study-q}~and~\ref{sec:large-study-q}.}
    \label{tab:hypotheses}
    \resizebox{\textwidth}{!}{
\begin{tabular}{@{}llllll@{}}
        \toprule
\textbf{Null ID}    & \textbf{Alt ID}    & \textbf{RQ}   & \textbf{MEM}   & \textbf{(Alternative) Hypothesis Text} &\textbf{Measurement Questions}
                                                                                                  \\ 
\midrule

\nullhypothesis{\hypoCheckUnderstand} & \althypothesis{\hypoCheckUnderstand} & \RQ{1} & \textbf{AE}
& Difference in check-for-understanding questions between the \ICS\ and \SEC\ students
&\texttt{SS-Q2, Q3, Q7, Q8, Q9, Q12, Q13, Q14, Q18}\\

\revised{\nullhypothesis{\hypoSecondADT}} & \revised{\althypothesis{\hypoSecondADT}} & \revised{\RQ{1}\& \RQ{4}}     & \textbf{AE}      & \revised{Difference in being able to successfully create ADTs}&\\

$\text{   }$\revised{\nullhypothesis{\hypoSecondADTsub}} & \revised{\althypothesis{\hypoSecondADTsub}} & \RQ{1}    & \textbf{AE}       & Difference in the successful creation of an ADT from a text description between the \ICS\ and \SEC\ students. &\texttt{LS-ADT2}\\

$\text{   }$\revised{\nullhypothesis{\hypoThirdADTsub}} &\revised{\althypothesis{\hypoThirdADTsub}} &  \revised{\RQ{1} \& \RQ{4}} & \textbf{AE} & \revised{Difference in the successful creation of an ADT from a self-selected scenario between the \ICS\ and \SEC\ students.}                 &\revised{\texttt{LS-ADT3}  }                        \\

\nullhypothesis{\hypoErrorAmount} &\althypothesis{\hypoErrorAmount} & \RQ{1} \& \RQ{4} & \textbf{AE} &  Difference in the number of errors made in ADT construction between the \ICS\ and \SEC\ students.            &\texttt{LS-ADT1},  \texttt{LS-ADT2}, \texttt{LS-ADT3}     \\

$\text{   }$\nullhypothesis{\hypoMultipleParent} &$\text{   }$\althypothesis{\hypoMultipleParent} & \RQ{1} \& \RQ{4} & \textbf{AE} & Difference in the number of multiple parent nodes used between the \ICS\ and \SEC\ students.  \\

$\text{   }$\nullhypothesis{\hypoMultipleRefinement} &$\text{   }$\althypothesis{\hypoMultipleRefinement} & \RQ{1} \& \RQ{4} & \textbf{AE} & Difference in the number of multiple refinement used between the \ICS\ and \SEC\ students.  \\

$\text{   }$\nullhypothesis{\hypoMulipleCountermeasure} &$\text{   }$\althypothesis{\hypoMulipleCountermeasure} & \RQ{1} \& \RQ{4} & \textbf{AE} & Difference in the number of multiple countermeasure nodes used between the \ICS\ and \SEC\ students.   \\

$\text{   }$\nullhypothesis{\hypoSingleChildNodes} &$\text{   }$\althypothesis{\hypoSingleChildNodes} & \RQ{1} \& \RQ{4} & \textbf{AE} & Difference in the number of single child nodes used between the \ICS\ and \SEC\ students.  \\

\nullhypothesis{\hypoSelfUnderstand} &\althypothesis{\hypoSelfUnderstand} & \RQ{2} & \textbf{PEOU} & Difference in the self-assessment between  \ICS\ and \SEC\ students.
&\texttt{LS-ADT1-L1}, \texttt{SS-Q5, Q10, Q15, Q19}\\

\nullhypothesis{\hypoCommunicationTool} &\althypothesis{\hypoCommunicationTool} & \RQ{2}     & \textbf{PU}      & Difference in the perception of usability of ADTs as a communication tool between the \ICS\ and \SEC\ students.             &\texttt{LS-ADT3-L3}           \\

\nullhypothesis{\hypoAnalysisTool} &\althypothesis{\hypoAnalysisTool} & \RQ{2}  &  \textbf{PU}       & Difference in the perception of usability of ADTs as an analysis tool between the \ICS\ and \SEC\ students.   &\texttt{LS-ADT1-L5}, \texttt{LS-ADT2-L2}, \texttt{LS-ADT3-L1}\\

\nullhypothesis{\hypoWrittenComparison} &\althypothesis{\hypoWrittenComparison} & \RQ{2}   &    \textbf{PEOU}    & Difference in the comparison of ADTs to a written description of attacks between the \ICS\ and \SEC\ students.
&\texttt{LS-ADT2-L1}, \texttt{LS-ADT3-L2}, \texttt{SS-Q6, Q11, Q16, Q20}    \\

\nullhypothesis{\hypoIntentionToUse} &\althypothesis{\hypoIntentionToUse} & \RQ{3}    & \textbf{ITU}       & Difference in the intention of students to use ADTs in the future between \ICS\ and \SEC\ students. & \texttt{LS-ADT3-W3}\revised{, \texttt{LS-ADT3-W5}} \\

\nullhypothesis{\hypoThirdADT} &\althypothesis{\hypoThirdADT} & \RQ{4}   & N/A        & Difference in the freely created ADTs of the \SEC\ and \ICS\ students.                &\texttt{LS-ADT3}                          \\
        \bottomrule
    \end{tabular}
    }
\end{table*}


\textbf{Protocol design.}
In the context of this work, we consider the technical background to be a background in computer science-related subjects. Our study was designed to measure how the difference in background affects the measured components.  Thus, we used a between-subjects design with two groups of students: one group with a strong computer science background, and another group with a very limited computer science background. Details about our participants are given further in Section~\ref{ssec:methodology-participants}. \revised{As common in such studies~\cite{labunetsSecurityRiskAssessment,lallieEmpiricalEvaluationEffectiveness2017,broccia_assessing_2024}, our participants first received training on the studied method (ADTs).}\ Both groups received the same lecture on ADTs given by the first author of this study, and afterward they participated in two identical study components: \emph{a small study}, which was an automatically assessed online quiz, and \emph{a large study} that involved a graded homework assignment. 

The two researchers involved in the study have several years of experience in teaching ADTs to diverse audiences (university students in Bachelor and Master programs, with and without a computer science background). This experience was instrumental in identifying the right questions and tasks for measuring the different components of interest. The study questionnaires were not pre-tested with the target student population as this was part of graded coursework and students who had seen the questions would have an unfair advantage; instead, the questionnaires were developed by taking advantage of the researchers' experience in teaching ADTs. The ethical considerations of our study are discussed in detail in Section~\ref{sec:ethics}.

\subsection{Study components}\label{sec:studycomponents}

\textbf{Small study.}
The small study contained 19 questions with each section starting with an image of an ADT with content and Likert questions for each ADT; each ADT was increasingly complex. This study focused on what information was received by looking at ADTs that were already created. All ADTs included in this assignment were taken from existing studies about ADTs~\cite{buldasAttributeEvaluationAttack2020,mauwRFIDCommunicationBlock,sunCyberAttackRisksAnalysis2018,kordyAttackdefenseTrees2014}. We selected such ADTs from the literature that a technical background would not be necessary to understand the attack scenario (i.e., without any specialist terms used for labels). All the questions and Likert statements can be found in Appendix~\ref{sec:small-study-q}.

Students did not receive a grade for completing the small study, but they were able to see the correct answers to the questions for self-evaluation immediately after completing the quiz. They were encouraged to do the quiz for their own learning, to ensure they understand ADTs as a concept, and as preparation for the larger homework assignment on ADTs and the final exam where ADTs were among the test questions.

This assignment focused on checking whether students are able to read ADTs and interpret them in the context of the studied theory (\revised{comprehension of the models}), as this was not a direct goal of the large study; although, \revised{as mentioned in Sec.~\ref{sec:choosing_method}, being able to interpret models correctly is necessary for the overall effectiveness of the method}. Further, an important purpose of the small study was to establish if the provided training was adequate.

\textbf{Large study.}
The large study was implemented as a take-home assignment, and students had four weeks to complete it at their own pace. This assignment was graded, contributing to the final course grade. Students were required to submit the assignment for the coursework, but they had to explicitly opt-in for participating in the study. We further discuss the ethical considerations of our study in Section~\ref{sec:ethics}.

This study consisted of three parts, with students creating attack trees in each part, under different conditions (from a set of components, from a given textual description, and for a self-selected scenario). Here, we aim to assess the more creative aspects of producing ADTs, which is the major motivation behind \RQ{4}. To our knowledge, this is also unique among TM studies, as to the best of our knowledge, nobody has yet examined creative aspects of threat model design. The list of questions from this study is available in Appendix~\ref{sec:large-study-q}.

\subsection{Data analysis}
\label{ssec:data-analysis}

Ultimately, since we start from the results by Lallie~\etal~\cite{lallieEmpiricalEvaluationEffectiveness2017}, we wish to find if there is a statistically significant difference between two independent treatment groups (those with a technical background and those with a very limited technical background). Much of our data is gathered through Likert questions, which result in ordinal data that cannot be normally distributed~\cite{verhulst2021best}, and for the remaining continuous data, we used the Shapiro-Wilk test to find that this data is not normally distributed~\cite{hanusz2016shapiro}. Our data also does not have equal variance according to Levene's test~\cite{levene1960robust}. Thus, we opt for the non-parametric Brunner-Munzel (BM) test~\cite{brunner2000nonparametric} that is robust in the unequal variance case~\cite{karch2021psychologists,fagerland2009wilcoxon}. As suggested by Labunets~\cite{labunets_no_2018}, when we do not find a statistically significant difference according to the BM test, we use the non-parametric Two One-Sided $t$-tests (TOST) to check for equivalence~\cite{schuirmann1981hypothesis}.

We correct for multiple tests using the Holm-Bonferroni (HB) correction method~\cite{holm1979simple} and adopt a significance threshold of $\alpha$=0.05, as is common practice in similar studies~\cite{labunets_no_2018,broccia_assessing_2024}.  In the remainder, we report the corrected $p$ values (denoted for short as $p * m$).

\subsection{Participants}
\label{ssec:methodology-participants}

\begin{figure}[t]
\includegraphics[width=\linewidth]{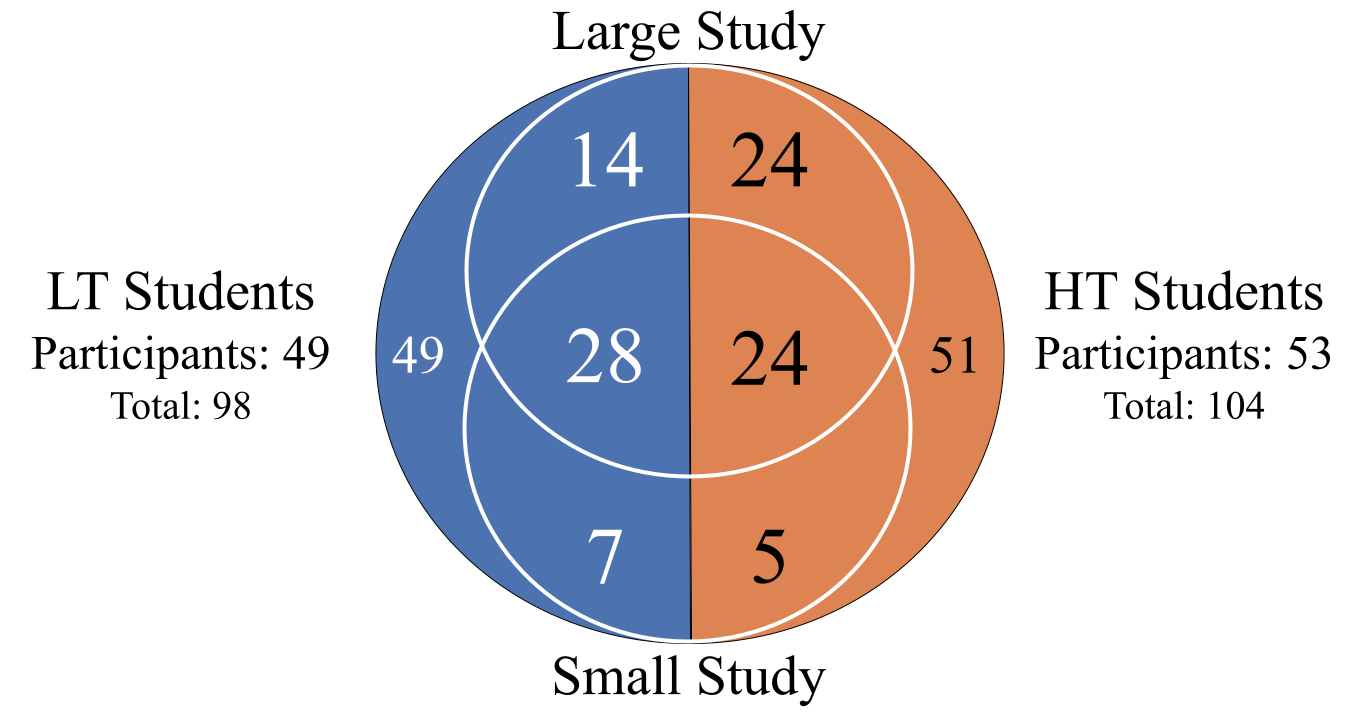}
    \caption{Distribution of participants across the treatment groups and studies.}
    \label{fig:participant-distribution}
\end{figure}

Participants in our study were undergraduate students at Leiden University (The Netherlands). The \ICS\ (\revised{Limited Technical}) students were predominately 3rd (final) year Bachelor students completing majors related to law, governance, and policy studies. The \ICS\ students were all a part of a minor focused on cyber security and governance. The \SEC\ (\revised{Highly Technical}) students were predominately 2nd year Bachelor students within the Computer Science Department.  Both groups of students were taking a major-appropriate Introduction to Cyber Security course, within which we ran our study. 

We consider that the \ICS\ students have a \revised{very limited technical}\ background and the \SEC\ students have a highly technical (computer science) background. This was confirmed with an optional demographic question asking participants how much programming experience they had. The \ICS\ participants had an average of 2.5 months of programming experience, which was the result of the \ICS\ students simultaneously taking a basic Python programming course (a component of the aforementioned minor)\footnote{\revised{This course is designed for students with zero programming experience. By the end of the course, students are expected to be able to write small (less than 30 lines) Python scripts that may integrate self-defined or imported functions and use objects.}}; in contrast, the \SEC\ students had an average of 3 years of programming experience. Additionally, according to their curriculum description, the \SEC\ students had two years of dedicated study in computer science, including courses on computer architecture, databases, linear algebra, algorithms, etc. These courses are not taken by the \ICS\ students.

Figure~\ref{fig:participant-distribution} provides the participant distribution between treatment groups in each experiment. There were a total of 49 \ICS\ (out of 98 taking the course) and 53 \SEC\ (out of 196 taking the course\footnote{In this group, it was possible to choose another assignment instead of ADTs, and 104 out of 196 students submitted the ADT assignment.}) consenting participants across the two studies. \revised{As the study was done in the educational context, we consider all submitted answers valid, even if part(s) of the questions were not answered. We reviewed all submissions and did not find evidence of invalid answers (e.g., participants who submitted intentionally wrong answers or answered randomly).}\ Table~\ref{tab:grades} shows a comparison of the final course grades (composed, in addition to the large study assignment, of an exam and several other assignment grades) of students in both treatment groups demonstrating that these groups are comparable to each other. While the grade analysis implies that stronger students self-selected to participate in the study, especially the optional small study, we can conclude that this is not different per students' background and study program.

\begin{table}
\caption{Comparison of the final course grades (out of 10) for participants and non-participants. \texttt{SS} stands for small study.}
\label{tab:grades}
\resizebox{\linewidth}{!}{
\begin{tabular}{lcccccccc}
\toprule
Type    & \multicolumn{2}{c}{Participant} & \multicolumn{2}{c}{Non-participant} & \multicolumn{2}{c}{BM Test} & TOST       & Effect Size                                     \\
& $n$   & mean grade     & $n$      & mean grade & statistic           & $p*m$        & $p*m$        & Cohen's $d$  \\
\midrule
\ICS\ (all) & 49 & 7.58 & 48 & 6.90 & -2.05 & 1.0 & \revised{1.0}& 0.70   \\
\SEC\ (all) & 53 & 7.52 & 48 & 6.39 & \revised{-2.12} & \revised{1.0} & \revised{0.79} & \revised{0.30}   \\
\midrule
\ICS\ (SS) & 35 & 7.64 & 63 & 7.23 & \revised{-1.678} & 1.0 & \revised{1.0} & 0.41  \\
\SEC\ (SS) & 29 & 7.54 & 72 & 6.68 & \revised{-1.193} & 1.0 & 1.0 & \revised{0.36}   \\
\midrule
\ICS & 49 & 7.58 &  &  & \multirow{ 2}{*}{\revised{1.384}} & \multirow{ 2}{*}{1.0} & \multirow{ 2}{*}{\revised{\textbf{0.037}}} & \multirow{ 2}{*}{\revised{0.11}}   \\
\SEC & 53 & 7.52  &  &   \\
\bottomrule
\end{tabular}
}
\end{table}

\subsection{Training}
\label{ssec:training}

\revised{As we mentioned, most of the empirical studies into the acceptability of security modeling methods provide training on the method as part of the study (see, e.g., \cite[Table\ 2.4]{labunetsSecurityRiskAssessment}, or previous studies of attack trees~\cite{lallieEmpiricalEvaluationEffectiveness2017,broccia_assessing_2024,broccia2025evaluating}.)}\ As our training, we gave a \revised{90 min.}\ lecture on threat modeling more broadly and ADTs in particular to both groups of students. The lecture covered an overview of threat modeling and a detailed introduction to ADTs with several examples. It also included an interactive component where students created their own ADTs, which were presented to the class as a whole with any issues or improvements discussed. A short description of the lecture and the slides are available in the provided data artifact~\cite{zenodo-dataset}. To ensure that both groups of students received a similar level of training, the slide deck and the lecturer were the same for both groups. 

The lecture to the \ICS\ students was given in October 2022, and the lecture to the \SEC\ students was given in February 2023. Both lectures were given in person without streaming or a recording being made. Attendance was encouraged but not required in both courses. There was an optional demographic question before the large study, which was answered by 29 participants in each treatment group (59\% of \ICS\ and 54\% of \SEC). Of these, 26 participants in each group indicated they attended the training lecture. The percentage range of training attendance for \ICS\ is 53\% - 93\% and the range for \SEC\ is 49\% - 87\%. Students had access to the detailed lecture slides while working on the study components at home.

\section{Study Results}\label{sec:results}
In this section, we present the study results per our main research questions \textbf{RQ1--RQ4}.

\begin{table*}[t!]
    \centering
    \caption{Check for understanding.}
    \label{tab:cfu-results}
    \resizebox{.68\textwidth}{!}{
    \begin{tabular}{@{}lcclclcccc@{}}

        \toprule
\textbf{Description}  & \textbf{Questions} & \multicolumn{2}{c}{\textbf{\ICS}} & \multicolumn{2}{c}{\textbf{\SEC\text{  }}} & \multicolumn{2}{c}{\textbf{BM test}} & \textbf{TOST} &     \textbf{Effect Size}             \\
&                    & $n$      & \% Correct        & $n$         & \% Correct    & statistic         & $p*m$   & $p*m$& Cohen's $d$                    \\ \midrule
Root nodes            & 1                  & 35       & 91.43             & 28          & 89.28         & -0.28                & 1.0          & \revised{\textbf{1.98e-17}} & 0.07 \\
Leaf nodes            & 3                  & 49       & 40.13             & 53          & 70.13         & \revised{3.73}                & \revised{\textbf{0.025}} &                 & 0.75 \\
Defense nodes         & 4                  & 49       & 51.19             & 53          & 61.13         & 1.25               & 1.0           & \revised{\textbf{2.65e-19}} & \revised{0.27} \\
Attack vectors        & 2                  & 35       & 34.29             & 28          & 44.64         & 0.84               & 1.0           & \revised{\textbf{6.19e-10}} & 0.24 \\
Levels of abstraction & 2                  & 34       & 26.47             & 27          & 37.04         &  \revised{1.36}               & 1.0          & \revised{\textbf{8.81e-17}} & 0.38 \\            \bottomrule
    \end{tabular}
    }
\end{table*}

\subsection{\RQ{1}: Effect of the background on \textbf{AE}}\label{sec:rq1}

\subsubsection{\hypothesis{\hypoCheckUnderstand}: Understanding ADT concepts}
\label{ssec:results-concept-understanding}

As mentioned in Section~\ref{ssec:training}, both \ICS\ and \SEC\ students received the same training in the form of a lecture. The lecture covered ADTs as a whole and delved into specific important concepts such as the types of nodes and refinements, levels of abstraction (LoA), and attack vectors. These concepts were addressed in detail during the lecture and practiced by students in small groups. We then tested the understanding of these concepts in the small study. 

\begin{figure}[h]
    \imgResize{
        \includegraphics[width=\linewidth]{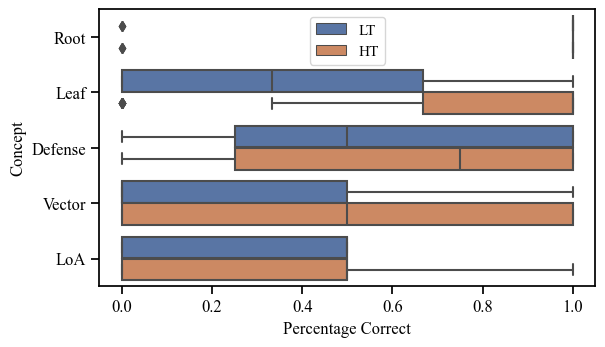}
    }
    \caption{Comparison of the average scores across check-for-understanding questions.}
    \label{img:cfu}
\end{figure}

In Table~\ref{tab:cfu-results} we see the aggregated responses to questions covering five chosen concepts related to ADTs. For each concept, Table~\ref{tab:cfu-results} presents the number of questions asked about each concept. We see the number of respondents from both groups as well as the average percentage of correct answers for each population and concept. Finally, we can see the statistics and $p * m$ values (with HB correction)  from the BM test. We can see that on four out of five concepts \ICS\ students scored, on average, somewhat worse than the \SEC\ students. However, only one of five topics (leaf nodes) has a statistically significant difference ($p * m <0.05$) between the two populations according to the BM test, and all topics show statistically significant results for equivalence according to TOST. We provide a visualization of this comparison in Figure~\ref{img:cfu}.

\highlight{\textbf{$H_1$}: We find evidence of equivalence between \ICS\ and \SEC\ students on understanding ADT concepts.}

\begin{figure}[t]
    \imgResize{
        \includegraphics[width=\linewidth]{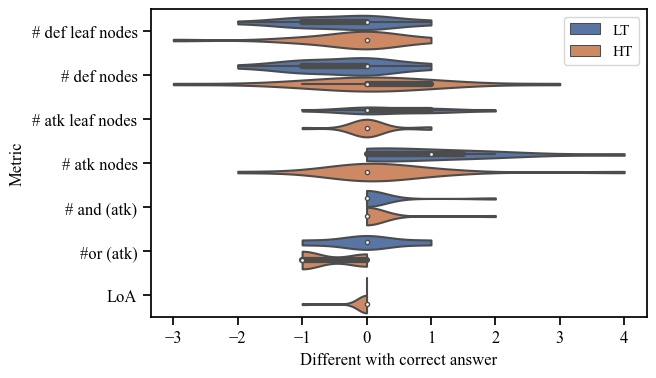}
    }
    \caption{Comparison on creating ADTs from a written description.}
    \label{fig:adt2-comparison}
\end{figure}

\subsubsection{\hypothesis{\hypoSecondADT}: Successfully creating ADTs}\label{ssec:results-h2}

\textbf{\hypothesis{\hypoSecondADTsub}: Creating ADTs from a written description.}
The second task of the large study was to create an ADT from a written description of an attack scenario. The written scenario was the result of reading out an existing ADT chosen by the research team into text. The students were tasked with reconstructing the original ADT from the text alone and were not told of the existence of the original ADT. They were specifically instructed to only include information from the scenario and to not introduce new information. From this task, we have 89 submitted ADTs \revised{(one participant did not submit an ADT for this task)}\ that are all nearly identical, as they are drawn from the same source material. Because of how the task was designed, we consider that this task has a correct answer. As such, we can compare the ADTs created by students to the original ADT to find where the participants deviated. 

Fifty-seven (57) ADTs (64\%) were identical to the original ADT according to the seven metrics we chose to measure the similarity of ADTs\footnote{To the best of our knowledge there is no established metric to measure distance or similarity between attack trees.}: the numbers of attack and defense nodes (we also separately count the number of attack and defense leaf nodes), the number of \OR\ and \AND\ refinements, and the number of levels of abstraction in a tree.  Of those identical ADTs, 26 (61.9\%) were provided by \ICS\ students, and 31 (64.6\%) were provided by \SEC\ students. \revised{The complete set of results is found in Table~\ref{h3h9-table}.}

Figure~\ref{fig:adt2-comparison} summarizes the 32 answers that deviated from the correct  ADT on at least one of the seven metrics. For example, if a student had one extra attack node, the figure would represent this answer as +1 in the ``\# atk nodes'' category. This figure shows that most students tended to make errors on only a few metrics, and produced results similar enough to the correct ADT. 
We see that only one \SEC\ student and no \ICS\ students made any errors regarding levels of abstraction (LoA); this could indicate that it is relatively easy for human participants to infer the different LoA from a textual description, and this holds for both participants with and without technical background.

\begin{table}[t!]s
    \centering
    \caption{Qualitative analysis of self-drawn ADTs.}
    \label{tab:results-qualitative}
    \revised{
        \resizebox{\linewidth}{!}{
    \begin{tabular}{lrlrlrlcccc}
        \toprule
    \textbf{Quality} & \multicolumn{2}{c}{\textbf{\shortstack{Largely\\Correct}}} & \multicolumn{2}{c}{\textbf{\shortstack{Neither}}} & \multicolumn{2}{c}{\textbf{\shortstack{Largely\\Incorrect}}}  & \multicolumn{2}{c}{\textbf{BM Test}} & \textbf{TOST} & \textbf{Effect Size} \\
     & \ICS & \SEC & \ICS & \SEC & \ICS & \SEC & statistic & $p*m$ & $p*m$ & Cohen's $d$ \\
     \midrule
Cohesive&21 & 22 & 15 & 19 & 4 & 6 & 0.46 & 1 & \textbf{8.21e-07} & 0.1\\
Clear&26 & 35 & 11 & 8 & 3 & 4 & -0.91 & 1 & \textbf{2.00e-07} & 0.15\\
Concise&24 & 24 & 16 & 20 & 0 & 3 & 1.20 & 1 & \textbf{6.96e-08} & 0.3\\
Complete&30 & 28 & 9 & 15 & 1 & 4 & 1.67 & 1 & \textbf{1.07e-06} & 0.35\\
    \bottomrule
    \end{tabular}
    }
    }
    \end{table}

\begin{table}
\centering
\caption{Results for hypotheses \revised{\hypothesis{\hypoSecondADTsub}}, \hypothesis{\hypoErrorAmount}, and \hypothesis{\hypoThirdADT}.}
\label{h3h9-table}
\resizebox{\linewidth}{!}{
\begin{tabular}{llllllc}
\toprule
\multicolumn{2}{l}{\textbf{Hypothesis}}         & \textbf{Component} & \multicolumn{2}{c}{\textbf{BM Test}} & \textbf{TOST} & \textbf{Effect Size}                             \\
            
&          && statistic             & \revised{$p*m$}     & \revised{$p*m$}              & Cohen's $d$      \\

\midrule
\revised{\multirow{7}{*}{\hypothesis{\hypoSecondADTsub}}} & & \revised{ADT2 defense leaf nodes} & \revised{0.656} & \revised{1.0} & \revised{\textbf{1.44e-11}}& \revised{0.07}\\
& & \revised{ADT2 defense nodes} & \revised{1.276} & \revised{1.0} & \revised{\textbf{4.88e-07}}& \revised{0.23}\\
& & \revised{ADT2 attack leaf nodes} & \revised{-1.435} & \revised{1.0} & \revised{\textbf{1.73e-16}}& \revised{0.33}\\
& & \revised{ADT2 attack nodes} & \revised{-1.727} & \revised{1.0} & \revised{\textbf{3.53e-04}}& \revised{0.31}\\
& & \revised{ADT2 AND (attack)} & \revised{-0.111} & \revised{1.0} & \revised{\textbf{3.77e-25}}& \revised{0.02}\\
& & \revised{ADT2 OR (attack)} & \revised{-2.496} & \revised{\textbf{.969}}& \revised{\textbf{5.15e-13}}& \revised{0.52}\\
& & \revised{ADT2 levels of abstraction} & \revised{-1.000} & \revised{1.0} & \revised{\textbf{5.57e-59}}& \revised{0.2}\\
\midrule\multirow{10}{*}{\hypothesis{\hypoErrorAmount}} &\multirow{2}{*}{\hypothesis{\hypoMultipleParent}}                      & ADT1 multi-parent nodes              & 0.32          & 1.0                   & \revised{\textbf{3.70e-13}} & 0.17 \\
&                  & ADT3 multi-parent nodes              & -0.64         & 1.0                   & \revised{\textbf{9.87e-17}} & 0.21  \\
\cmidrule{2-7}&\multirow{2}{*}{\hypothesis{\hypoMultipleRefinement}}                    & ADT1 multi refinement                & -0.53         & 1.0                   & \revised{\textbf{8.23e-15}} & 0.03  \\
&                    & ADT3 multi refinement                & -0.13         & 1.0                   & \revised{\textbf{3.54e-26}} & 0.04 \\
\cmidrule{2-7}&    \multirow{3}{*}{\hypothesis{\hypoMulipleCountermeasure}}                & ADT1 multi countermeasure            & 1.76          & 1.0                   & \revised{\textbf{4.59e-12}} & 0.38 \\
&                    & ADT2 multi countermeasure            & 0.46          & 1.0                   & \revised{\textbf{9.86e-22}} & \revised{0.02} \\
&                    & ADT3 multi countermeasure            & -0.96         & 1.0                   & \revised{\textbf{0.035} }    & \revised{0.1}  \\
\cmidrule{2-7}& \multirow{3}{*}{\hypothesis{\hypoSingleChildNodes}}                   & ADT1 single child  (attack)              & -4.68         & \revised{\textbf{8.66e-04}}      &               & 0.79  \\
&                    & ADT2 single child  (attack)              & 1.50          & 1.0                   & \revised{\textbf{1.01e-10}} & 0.19 \\
&                    & ADT3 single child  (attack)              & -2.66         &  \revised{0.641}                & \revised{1.0}              & 0.47  \\

\midrule
\multirow{7}{*}{\hypothesis{\hypoThirdADT}}     &                    & ADT3 defense leaf nodes               & -0.35         & 1.0                   & 1.0               & 0.10 \\
&                    & ADT3 defense nodes                    & -0.31         & 1.0                   & 1.0               & 0.13 \\
&                    & ADT3 attack leaf nodes               & 0.83          & 1.0                   & 1.0               & 0.36 \\
&                    & ADT3 attack nodes                    & 0.19          & 1.0                   & 1.0               & 0.24 \\
&                    & ADT3 AND  (attack)                    & 1.24          & 1.0                   & 1.0               & 0.34 \\
&                    & ADT3 OR  (attack)                      & 0.40          & 1.0                   & 1.0               & 0.23 \\
&                    & ADT3 levels of abstraction                             & -0.53         & 1.0                   & \revised{0.105}     & 0.10  \\
&                    & \revised{ADT3 and:or ratio}                          & \revised{0.19}         & \revised{1.0}                   & \revised{\textbf{1.95e-03}}     & \revised{0.11}  \\
\bottomrule
\end{tabular}
}

\end{table}


\textbf{\hypothesis{\hypoThirdADTsub}: Creating ADTs for a self-selected scenario.}
It is important to assess whether the participants are able to produce high-quality ADTs to represent a diverse set of attack scenarios. In total, there were 88 ADTs (two participants did not submit an ADT for this task) drawn for the task where students had to model their own scenarios. 

We qualitatively evaluated the ADTs designed for self-selected scenarios (we call them \emph{self-drawn ADTs}) based on four criteria: how meaningful are the refinements (\emph{cohesiveness}), how clear are the labels (\emph{clarity}), how relevant are the suggested attack components and whether there are any excessive steps (\emph{conciseness}), and how complete are the scenarios (\emph{completeness}). These qualities were selected to represent together a quality evaluation of the designed models. 

 The evaluation was done by two researchers experienced in attack trees and cybersecurity. First, the researchers designed together a rubric to evaluate ADTs based on these four criteria. The rubric was adjusted and calibrated in two iterations, when the researchers would first independently evaluate a set of randomly selected ADTs from both \ICS\ and \SEC\ participants and then jointly discuss the results. In the second iteration, the two researchers independently assessed all considered trees in the same way (reaching an agreement).  This final rubric used to evaluate the ADTs according to these criteria is available in the provided data artifact~\cite{zenodo-dataset}. The principal researcher then evaluated the whole set of ADTs based on the final rubric. The results of the evaluation according to this rubric can be found in Table~\ref{tab:results-qualitative}, which shows that there is statistically significant equivalence between the groups on all four criteria.

\highlight{\textbf{$H_2$}: We find no significant evidence of a difference between \ICS\ and \SEC\ students on effectively creating ADTs.}

\subsubsection{\hypothesis{\hypoErrorAmount}: Common errors when designing ADTs}
\label{ssec:results-common-errors}

Another metric we used to compare the two populations of students is the common mistakes they made while creating ADTs. After manually checking all 180 received ADT images, we identified four common types of mistakes described below.

\textbf{\hypothesis{\hypoMultipleParent}: Multi-parent nodes.} These describe nodes that have more than one parent. ADT construction rules (syntax) allow only a single parent for every node~\cite{kordyFoundationsAttackDefense2011}. For each node that had more than one parent, we counted that node as an error. If a node had more than two parents, the node was still counted only once.

\textbf{\hypothesis{\hypoMultipleRefinement}: Multi-refinement nodes} These are nodes that have children with multiple refinement relationships. ADT construction rules allow for one refinement per node, in our case either \AND\ or \OR~\cite{kordyFoundationsAttackDefense2011}. Some students would have two child nodes in an \AND\ relationship, and then a third or fourth child node that was not included in the \AND. This was expressed by the \AND\ arc not extending to the connecting edge of these other children. It was clear to us, also based on the node labels, that some children were in an \AND\ relationship, while the remainder was in an \OR\ relationship. We counted each node with multiple refinements regardless of the number of children that node had.

\textbf{\hypothesis{\hypoMulipleCountermeasure}: Multi-countermeasure attack nodes.} These are attack nodes that have multiple countermeasures. ADT construction rules only allow for one countermeasure child per node~\cite{kordyFoundationsAttackDefense2011}. If multiple countermeasures are possible, there should first be an intermediate defense node with the single countermeasure edge, and then the multiple countermeasures can be added to the intermediate node in either \AND\ or \OR\ relationship. We counted each time an attack node had more than one countermeasure, regardless of the number of countermeasures attached to that node. 

\textbf{\hypothesis{\hypoSingleChildNodes}: Single-child nodes.} These are nodes that had only one child node. This type of error is unlike the previous three in that it is not a semantic error. Semantically, there is no issue with having a single child, with multiple semantic representations of ADTs allowing a single child node~\cite{mauwFoundationsAttackTrees2006,kordyFoundationsAttackDefense2011}. A single child node can be shown to be equivalent in both \AND\ and \OR\ refinements, thus technically we can admit attack trees with such refinements as valid.
The primary reason for single-child nodes to be included in this section is students were explicitly instructed to avoid using single-child nodes, as the syntactic ADT definition requires that each refined node has at least two children of the same type in either \AND\ or \OR\ relationship, and if only one child is needed, it can be absorbed in the parent node itself. We acknowledge that this argument is flawed for practical reasons, as single child nodes may be necessary to cognitively help the analysts to consider different sub-scenarios and keep the levels of abstraction of a tree consistent across different branches. However, levels of abstraction and the cognitive needs of the analysts were not a focus of our research, while the use of ADTs in a syntactically correct manner was a focus; thus, we have elected to consider single child nodes as an error.


\begin{figure}[t]
    \imgResize{
        \includegraphics[width=0.8\linewidth]{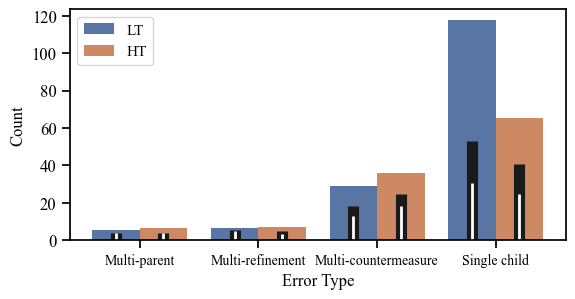}
    }
    \caption{Comparison of of the amount of semantic errors made by \ICS\ and \SEC\ students.}
    \label{fig:error-amounts}
\end{figure}

\begin{figure}[t]
    \imgResize{
        \includegraphics[width=0.8\linewidth]{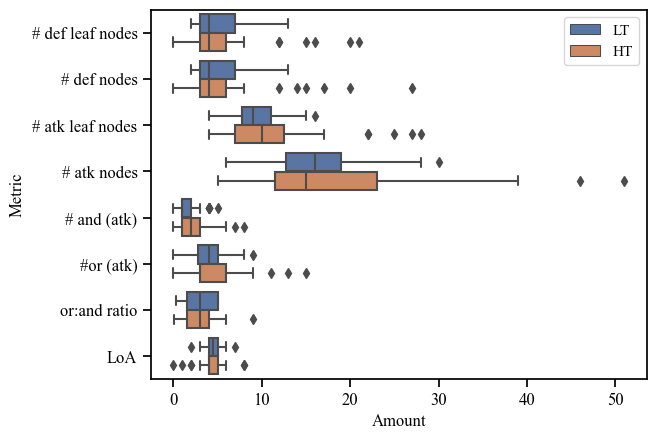}
    }
    \caption{Comparison of \ICS\ and \SEC\ students' self-drawn ADTs on \revised{quantitative}\ metrics. }.
    \label{fig:adt3-comparison}
\end{figure}

\textbf{Analysis of common errors.}
Figure~\ref{fig:error-amounts} shows the total number of errors present in the ADTs of both \ICS\ and \SEC\ students. The colored bars show the total error count; if a student made an error three times on the same ADT, then this would be counted three times in the total error count. By contrast, the small black bar inside each colored bar shows the total number of ADTs that have errors in them (the large study consisted of three separate ADTs). The small white bar within the black bar shows the total number of students who made these errors. If the height of the colored and black bars is similar, it indicates that the number of errors present per ADT is closer to 1. If the height of the white and black bar is similar, this indicates that students only made this mistake on one of their three ADTs; a significant height difference here indicates that some students made this mistake on more than one ADT.

In Figure~\ref{fig:error-amounts}, we see that multi-refinement and multi-countermeasure errors are made very infrequently at very similar rates between \ICS\ and \SEC\ students. For the single-child error (\hypothesis{\hypoSingleChildNodes}), we see that a similar number of students made these errors across similar numbers of ADTs; however, \ICS\ students made this error nearly twice as many times as \SEC\ students (this difference is statistically significant in ADT1 according to the BM test). The results of our testing can be found in Table~\ref{h3h9-table}.  Across the other errors \hypothesis{\hypoMultipleParent}, \hypothesis{\hypoMultipleRefinement}, and \hypothesis{\hypoMulipleCountermeasure}, there is no statistically significant difference between \ICS\ and \SEC\ students, but there is statistically significant equivalence according to TOST.

\highlight{\textbf{$H_3$}: We find a significant difference between the groups with respect to single-child nodes. We see evidence of groups' equivalence for all other types of errors. Overall, we find little evidence of a difference and significant evidence of equivalence between \ICS\ and \SEC\ on common errors.}

\textbf{Conclusions on the actual effectiveness of ADTs.}
We can conclude that, while we observed a statistically significant difference between the treatment groups for the two types of errors we considered, the majority of the other tested components of the actual effectiveness show the absence of a statistically significant difference between groups' performances. On some measured components, like the quality of self-drawn ADTs, the two treatment groups show statistically significant equivalent behavior. Overall, while both groups show the same lack of understanding of some aspects of ADTs, both groups have demonstrated sufficient mastery of the topic at a similar rate, allowing us to conclude that the actual effectiveness of ADTs is high for both groups.

\highlight{\RQ{1}: Actual effectiveness of ADTs is high for both groups and does not appear to be affected by technical background.}


\subsection{\RQ{2}: Effect of the background on \textbf{PU} and \textbf{PEOU}}\label{sec:rq2}

\subsubsection{Perceived Ease of Use (\textbf{PEOU})}

\textbf{\hypothesis{\hypoSelfUnderstand}: Self-assessment of understanding.}
Alongside the check-for-understanding questions we discussed in Section~\ref{ssec:results-concept-understanding}, we asked students if they found a given ADT easy to understand. For the small study, we asked if the provided ADT was easy to understand, and for the large study, we asked if the structure of ADTs was easy to understand. These questions were all in service of the same goal: assessing how students perceived their own understanding of ADTs.

\begin{figure}[t]
    \imgResize{
        \includegraphics[width=\linewidth]{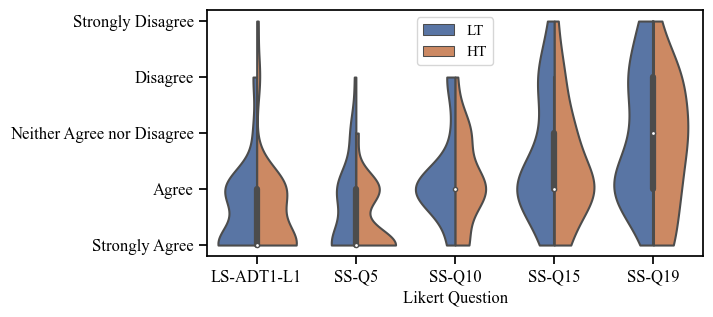}
    }
    \caption{Comparison of \ICS\ and \SEC\ students on responses to questions self-assessing their understanding of ADTs. The ADTs used in the questions are referenced in Appendices~\ref{sec:small-study-q} and \ref{sec:large-study-q}.}
    \label{img:likert-understanding}
\end{figure}

In general, \ICS\ and \SEC\ students both assessed their understanding similarly (see Figure~\ref{img:likert-understanding}). With the small study questions (labeled \texttt{SS-Q\#}), the students reported a steady decrease in their confidence in understanding. This is to be expected since, as we describe in Sec.~\ref{sec:studycomponents}, there were four ADTs with increasing complexity. The same question was asked about each ADT, and students were less confident with more complex trees.

In Table~\ref{tab:likert}, we can see that none of the understanding Likert questions shows any statistically significant difference between the groups according to the BM test (and some of the questions demonstrate significant equivalence of the groups). 

\highlight{\textbf{$H_4$}: We find evidence of equivalence between \ICS\ and \SEC\ students on self-assessment of understanding.}

\begin{table*}[t!]
\centering
\caption{Table showing the statistics and analysis of answers to Likert questions per hypothesis and treatment group.}
\label{tab:likert}
\resizebox{0.88\textwidth}{!}{
\begin{tabular}{@{}llrlrlrlrlrlrllllc@{}} 
\toprule
\textbf{Hypothesis}           & \textbf{Question}\footnotemark   & \multicolumn{2}{c}{\textbf{Str. Agree}} & \multicolumn{2}{c}{\textbf{Agree}} & \multicolumn{2}{c}{\textbf{Neither}} & \multicolumn{2}{c}{\textbf{Disagree}} & \multicolumn{2}{c}{\textbf{Str. Disagree}} & \multicolumn{2}{c}{\textbf{Average}} & \multicolumn{2}{c}{\textbf{BM test}}      & \textbf{TOST}& \textbf{Effect Size} \\&                 &     \ICS            &      \SEC           &     \ICS            &     \SEC            &     \ICS            &    \SEC   &    \ICS   &   \SEC    &   \ICS    &   \SEC    & \ICS   &     \SEC   &statistic&$p*m$         &$p*m$  &Cohen's $d$        \\ \midrule
\multirow{5}{*}{\shortstack[l]{Understanding\\(\hypothesis{\hypoSelfUnderstand})}}       

& \texttt{LS-ADT1-L1} & 21           & 26  & 17    & 19     & 2               & 0     & 2  & 2  & 0 & 1 & 1.64 & 1.6 &  -0.46& 1.0  &\revised{\textbf{1.92e-05}}&0.05\\
& \texttt{SS-Q5}      & 17           & 17  & 13    & 10     & 4               & 1     & 1  & 0  & 0 & 0 & 1.69 & 1.43 & -1.25& 1.0   &\revised{\textbf{4.30e-03}}&0.37\\
& \texttt{SS-Q10}     & 4            & 6   & 24    & 14     & 2               & 7     & 5  & 1  & 0 & 0 & 2.23 & 2.11 & -0.28& 1.0    &\revised{\textbf{3.26e-03}}&0.15\\
& \texttt{SS-Q15}     & 4            & 4   & 17    & 13     & 4               & 6     & 7  & 3  & 2 & 1 & 2.59 & 2.41 & -0.50& 1.0   &\revised{0.162}&0.17\\
& \texttt{SS-Q19}     & 2            & 5   & 14    & 6      & 5               & 8     & 8  & 7  & 4 & 1 & 2.94 & 2.74 & -0.49& 1.0    & \revised{0.395}&0.17\\\midrule

Communication (\hypothesis{\hypoCommunicationTool}) 

& \texttt{LS-ADT3-L3} & 24           & 28  & 11    & 14     & 1               & 3     & 2  & 3  & 2 & 0 & 1.68 & 1.6  & 0.06& 1.0   &\revised{\textbf{1.14e-03}}&0.08\\\midrule

\multirow{3}{*}{\shortstack[l]{Analysis\\(\hypothesis{\hypoAnalysisTool})}}            

& \texttt{LS-ADT1-L5} & 19           & 18  & 12    & 17     & 5               & 5     & 5  & 5  & 1 & 2 & 1.98 & 2.06 & 0.45& 1.0   & \revised{\textbf{0.013}}&0.07\\
& \texttt{LS-ADT2-L2} & 28           & 13  & 8     & 22     & 3               & 3     & 0  & 5  & 3 & 5 & 1.62 & 2.31 & 3.60 & \revised{\textbf{0.041}}  & & 0.57\\
& \texttt{LS-ADT3-L1} & 21           & 16  & 16    & 18     & 2               & 5     & 2  & 5  & 1 & 4 & 1.71 & 2.23 & 2.14 & \revised{1.0} &   \revised{1.0} &0.46\\ \midrule

\multirow{6}{*}{\shortstack[l]{Written\\description\\(\hypothesis{\hypoWrittenComparison})}} 

& \texttt{LS-ADT2-L1} & 18           & 24  & 12    & 12     & 5               & 6     & 3  & 3  & 4 & 3 & 2.12 & 1.94 & -0.68 & 1.0  &\revised{0.101}&0.14\\
& \texttt{LS-ADT3-L2} & 6            & 7   & 11    & 14     & 7               & 3     & 16 & 21 & 2 & 3 & 2.93 & 2.98  & 0.26& 1.0  &\revised{\textbf{0.018}}&0.04\\
& \texttt{SS-Q6}      & 11           & 13  & 13    & 13     & 7               & 0     & 4  & 1  & 0 & 1 & 2.11 & 1.71 & -1.89 & 1.0   &\revised{0.595}&0.41\\
& \texttt{SS-Q11}     & 10           & 7   & 14    & 16     & 6               & 3     & 5  & 1  & 0 & 1 & 2.17 & 2.04 & -0.51& 1.0   &\revised{\textbf{0.034}}&0.14\\
& \texttt{SS-Q16 }    & 12           & 7   & 10    & 14     & 4               & 2     & 8  & 2  & 0 & 2 & 2.24 & 2.19 & -0.05& 1.0   &\revised{0.092}&0.04\\
& \texttt{SS-Q20}     & 11           & 10  & 12    & 7      & 3               & 5     & 5  & 3  & 2 & 2 & 2.24 & 2.26 & 0.01 & \revised{0.994}     &\revised{0.146}&0.01\\
\bottomrule
\end{tabular}
}
\end{table*}



\begin{figure}[t]
    \imgResize{
        \includegraphics[width=\linewidth]{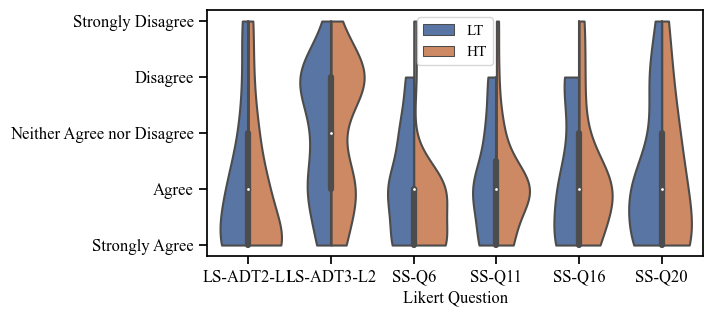}
    }
    \caption{Responses to questions concerning the preference of ADTs to a written description.}
    \label{fig:likert-written}
\end{figure}

\textbf{\hypothesis{\hypoWrittenComparison}: Written description preference.}
We asked students across every ADT model in the small study and across the final two ADTs in the large study if they prefer ADTs to a written description of an attack scenario. In all questions save one, there was no written description provided; students were asked if their preference was for an ADT that was either presented or to an ADT they had drawn, without an alternative written text about the scenario present (there is one exception to this: the task on building an ADT in the large study where students converted a textual attack scenario description to an ADT). The responses for both \ICS\ and \SEC\ students were similar: Table~\ref{tab:likert} shows that there is a statistically significant equivalence between \ICS\ and \SEC\ for questions in the written description category. This is also demonstrated by Figure~\ref{fig:likert-written}.

\highlight{\textbf{$H_7$}: We find evidence of equivalence between \ICS\ and \SEC\ students on preference of ADTs to a written description.}

\subsubsection{Perceived Usefulness (\textbf{PU})}

\begin{figure}[t]
    \imgResize{
        \includegraphics[width=\linewidth]{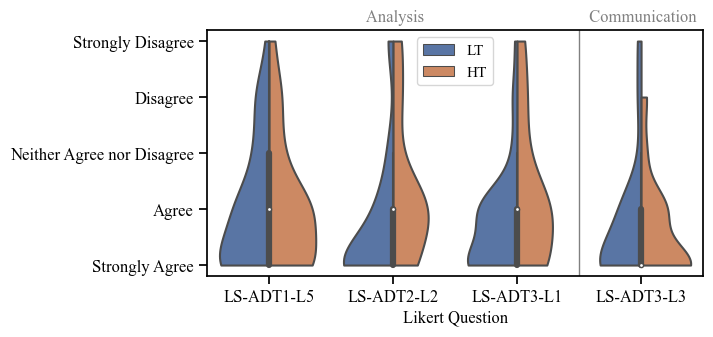}
    }
    \caption{Replies concerning ADTs as a means of analysis and communication.}
    \label{fig:means-of-commanalysis}
\end{figure}

\textbf{\hypothesis{\hypoCommunicationTool}\&\hypothesis{\hypoAnalysisTool}: ADTs as a means of analysis and communication.}
We asked three questions about how students perceived ADTs as a means of analysis and one question about how they perceived ADTs as a means of communication. The data shape of responses can be seen in Figure~\ref{fig:means-of-commanalysis}.

We have more detailed information in Table~\ref{tab:likert}, where we see strong equivalence between \ICS\ and \SEC\ students when considering ADTs as a means of communication. Both groups overwhelmingly agree that ADTs are useful as a tool for communicating attack scenarios. We see more agreement than disagreement about ADTs as a means of analysis, however, it is not as strong as the agreement we see for ADTs as a means of communication. Additionally, we see a statistically significant difference on two of the three questions concerning ADTs as a means of analysis. On these two questions, the \ICS\ students agreed more than the \SEC\ students that ADTs are a useful tool for analysis, with moderate effect sizes (see the Cohen's $d$ values in Table~\ref{tab:likert}).

\highlight{\textbf{$H_5\&H_6$}: \ICS\ and \SEC\ students equally perceive ADTs to be useful as a means of communication, but we find some evidence of a difference in their perceptions of ADTs as a means of analysis.}

\textbf{Conclusions on perceptions of ADTs.}
Overall, we find that the treatment groups largely perceived ADTs to be useful and easy to use (thus, the perceived efficacy is high). \textbf{PEOU} is statistically significantly equivalent in both groups, while \textbf{PU}, while similar, is not equivalent, and is significantly diverging on one measured aspect (ADTs perceived as a useful means of analysis when designing a model from a textual description). 

The only aspect for which we have found a statistically significant difference between the populations revolved around the Likert question concerning ADTs as a means of analysis. One interpretation of this result could be that \ICS\ students were introduced to a novel means of organizing information (in the tree structure), which would aid in analysis. In contrast, \SEC\ students should have seen tree structures in their previous coursework, which would lead to ADTs not introducing a new means of organizing information. 
This hypothesis would need further study in order to be tested.

\highlight{\RQ{2}: We find little evidence that the perceived efficacy of ADTs is affected by technical background. The only hypothesis \hypothesis{\hypoAnalysisTool} for which we have observed a statistically significant difference affects the perception of ADTs in a specific context only, as a means of analysis. The perceived efficacy of ADTs is high for both groups.}


\subsection{\RQ{3}: Effect of the background on \textbf{ITU}}\label{sec:rq3}

\textbf{\hypothesis{\hypoIntentionToUse}: Intention to use.}
We asked two open questions relevant to this hypothesis: \texttt{LS-ADT3-W3} asked the participants if they believe ADTs have a place in the cybersecurity field, and if so, where, while \texttt{LS-ADT3-W5} asked the students if they would like to see ADTs again. To analyze these questions, we applied a simple coding. If students responded in the affirmative, we applied a value of $1$ to the code ``Yes''. If the student replied in the negative, we applied a value of $0$, and if the student replied in a manner that was open to interpretation, we applied a value of $0.5$. We followed this structure for the other codes. The ``Communication'' code refers to a response describing the utility of ADTs as a means of communication and the ``Analysis'' code refers to a response describing the utility of ADTs as a means of analysis. These codes are not mutually exclusive, as many responses were coded as neither or both. In this way, we obtain a quantitative evaluation of a qualitative question. \revised{The coding guidelines were developed by the two researchers together, and several randomly selected answers from each category were evaluated independently to verify that the assessment aligns. After the establishment of the guidelines, the coding was done by a single coder (the first author of this work).  }

Table~\ref{tab:coded-future-use} contains the \ICS\ and \SEC\ averages of these codes.
We can see that there is a statistically significant equivalence between the responses. Additionally, we see that both \ICS\ and \SEC\ students strongly agreed that ADTs have a place in the cybersecurity industry, and fairly strongly agreed that they would like to see ADTs again in the future.

\highlight{\textbf{$H_8$}: We find evidence of equivalence between the treatment groups on intention to use ADTs.}

\textbf{Conclusions on intention to use ADTs.}

\highlight{\RQ{3}: The intention to use ADTs is high for both groups and is not affected by technical background.}


\subsection{\RQ{4}: Effect of the background on creative aspects of ADT design}\label{sec:rq4}

While the equivalence of two ADTs can be assessed based on a chosen semantics~\cite{mauwFoundationsAttackTrees2006}, to the best of our knowledge, ADT comparison and metrics of distance between two ADTs have not yet been investigated in the literature. Thus, we opted to compare the self-drawn ADTs based on several \revised{quantitative and qualitative}\ metrics.

\textbf{\hypothesis{\hypoThirdADT}: Self-drawn ADT comparison.}
The third task in the large study required the participants to design an ADT for their scenario of choice. As we mentioned in Sec.~\ref{sec:studycomponents}, we intentionally did not give any indication of the acceptable size for the tree, as we wanted to assess what differences, if any, would appear between ADTs drawn by \ICS\ and \SEC\ students when there are no priming restrictions, thereby evaluating the creative component.

We \revised{quantitatively}\ assessed the ADTs on 8 metrics: the total number of attack and defense nodes, the number of attack and defense leaf nodes, the number of \OR\ and \AND\ refinements, the ratio of \OR\ to \AND\ refinements, and the levels of abstraction. For these criteria, we define a leaf node as any node that does not have children of the same type. Thus, a node that only has a countermeasure edge would also be defined as a leaf node. We define levels of abstraction to be the greatest depth in the tree, not including countermeasures.

We compared \ICS\ and \SEC\ students' answers on these eight metrics using the BM test and found that there is no statistically significant difference between the ADTs drawn by \ICS\ and \SEC\ students on any metric. The results of our testing can be found in Table~\ref{h3h9-table}. Overall, we find the ADTs drawn by these two groups of students to be remarkably similar (though not equivalent in a statistically significant way).

\begin{figure}[h]
        \imgResize{
            \includegraphics[width=0.68\linewidth]{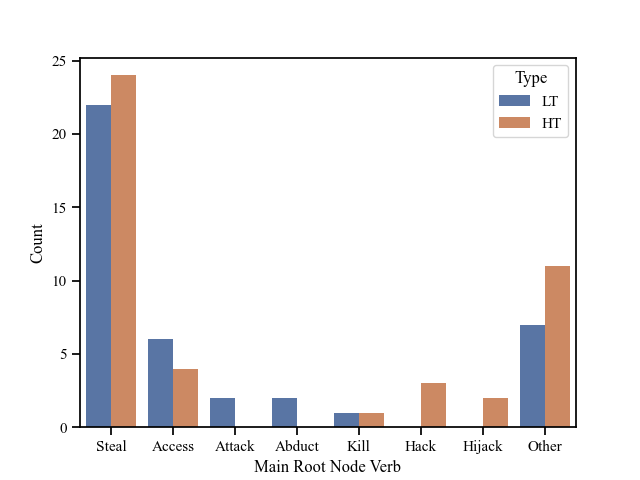}
        }
        \caption{\revised{Comparison of main verbs in the root nodes of the self-drawn ADTs.}}
        \label{fig:root-verbs}
    \end{figure}

\revised{We qualitatively evaluated the trees using two methods. Besides the quality evaluation results reported in Sec.~\ref{ssec:results-h2} that show that both groups designed ADTs with equivalent quality, we processed the labels of the root nodes, taking the main verb from each label (when present) and standardizing these (for example, ``steal'' and ``rob'' were considered equivalent in meaning). In Figure~\ref{fig:root-verbs}, we can see the prevalence of verbs across the two groups for all verbs that were present in at least two ADTs. While there are some differences in the verbs, as with the quality analysis, overall, the verbs used in the root nodes are similar between the groups.}

\highlight{\textbf{$H_9$}: We find no evidence of difference between the treatment groups on self-drawn ADTs.}

\begin{table}[t!]
    \caption{Coded responses to written questions concerning the future use of ADTs.}
    \label{tab:coded-future-use}
    \resizebox{.48\textwidth}{!}{
        \begin{tabular}{@{}llllllll@{}}
            \toprule
            \textbf{Question} & \textbf{Code} & \multicolumn{2}{c}{\textbf{Average}} & \multicolumn{2}{c}{\textbf{BM Test}} & \textbf{TOST} &\textbf{Effect Size}                 \\
     &               & \ICS        & \SEC               & statistic           & $p*m$   & $p*m$  &Cohen's $d$  \\\midrule
     \texttt{LS-ADT3-W3}        & Yes           & 0.92        & 0.96               & \revised{0.90}       & 1.0   & \revised{\textbf{3.99e-40}} &\revised{0.23}\\
     \texttt{LS-ADT3-W3}        & Communication & 0.61        & 0.52               & \revised{-0.91}       & 1.0 & \revised{\textbf{3.57e-13}} &\revised{0.19}\\
     \texttt{LS-ADT3-W3}        & Analysis      & 0.44        & 0.45               & \revised{0.12}       & 1.0   & \revised{\textbf{3.98e-15}} &\revised{0.02}\\
     \texttt{LS-ADT3-W5}        & Yes           & 0.85        & 0.73               & \revised{-1.91}       & 1.0 & \revised{\textbf{1.37e-17}} &0.31\\ \bottomrule
        \end{tabular}
    }
\end{table}

\textbf{Conclusions on creative expression with ADTs.}
If ADTs were understood and used differently, we would expect to see a statistically significant difference in the ADTs created by the two groups on some qualitative or quantitative metric. As we cannot see a significant and material difference, this supports our conclusion that the technical background does not impact how ADTs are created.

\highlight{\RQ{4}: The creative component in creating ADTs is not affected by technical background.}

\section{Discussion}\label{sec:discussion}

Our results show that both participants with a highly technical background and a very limited technical background find ADTs acceptable. Moreover, they find it acceptable in an equal way: for most of the concepts we measured, both treatment groups have shown equivalent behavior and perceptions. They also use ADTs creatively in a similar way, designing models of very similar size and quality. These findings confirm the belief in the security community that attack trees are accessible and easy to learn~\cite{reversinglabs2024attacktrees}. 

Our research sought to establish if the technical background is a potential factor in the adoption of ADTs and, specifically, if the participants with a very limited technical background would be disadvantaged in using ADTs. The cyber security industry consists of people with widely varying backgrounds~\cite{andersonSecurityEngineeringGuide2020}. In particular, TM is done by people with diverse skillsets and objectives~\cite{shostack2008experiences,verreydt2024threat}. If a technical background were to impact the acceptability of ADTs, then this could be a reason for not recommending them to be used.

Lallie~\etal\ found that there was a difference between participants with and without a computer science background when using both fault trees and attack graphs in a similar study design to ours~\cite{lallieEmpiricalEvaluationEffectiveness2017}. This result indicated that TM stakeholders who do not possess a highly technical background (e.g., managers) might potentially be disadvantaged if the team uses attack trees for threat modeling. However, it is reasonable to expect that people involved in TM, even managers, might possess at least a limited technical background as they are exposed to software development and/or IT security risk management activities. Our study concludes that ADTs are highly acceptable for such TM stakeholders and do not disadvantage them compared to threat modelers with a highly technical background.

We believe the difference in the results between our study and \cite{lallieEmpiricalEvaluationEffectiveness2017} to be due to two major differences in the study design and methodology. First, we intentionally used ADT examples that are equally accessible to all participants, attempting to remove any specifically technical jargon from our study questions. For the small study, all of our examples were pulled from papers on ADTs and we specifically looked for ADTs without complex technical labels, i.e., accessible to people with diverse backgrounds. This approach was inspired by Lallie et al.~\cite{lallieEmpiricalEvaluationEffectiveness2017} who used fault trees from previous works. However, two of the fault trees they used are arguably difficult to understand to a layperson, using terms such as ``\texttt{sshd\_bof(1,2)}'', which might be more accessible to someone with a computer science background. As such, their finding that those with a computer science background can use these models more effectively may speak more to the comprehensibility of the language used in their study. While subsequent studies in the attack trees context are required to test this,  previous research has shown that technical language does affect comprehension: e.g., Bravo-Lillo et al.~\cite{bravo2010bridging} have shown that technical terms complicate comprehension of security warnings for non-expert users, compared to security experts. In the TM context, Ingalsbe et al.~\cite{ingalsbe2008threat} explicitly mention that the vocabulary of threat modeling is IT-biased, impeding communication with internal business customers, while Verreydt et al.~\cite{verreydt2024threat} also acknowledge the TM challenges related to communication and the used language.

One important conclusion that we can draw from our study is that short training is sufficient for making ADTs equally acceptable for users with high and limited technical backgrounds. Threat modeling method training is an established practice in organizations~\cite{verreydt2024threat}, and it can be recommended to improve the outcomes and facilitate the process~\cite{cruzes2018challenges,stevens2018battle}. To help implement training on ADTs in organizations, we share the slides of our training lecture along with a detailed description in our supplementary data material~\cite{zenodo-dataset}.

Another relevant observation that we can make from the analysis of the related literature (Sec.~\ref{sec:empirical_studies}) is that there are no established protocols for empirical studies of TM methods. While the studies frequently follow reputable frameworks like TAM and MEM, the operationalization of the frameworks' constructs differs a lot. One of the reasons behind this might be the diversity of TM methods themselves. Still, it would be useful to systematize the experiences reported so far and develop guidelines for executing such studies.

\section{Limitations}\label{sec:limitations}
Our study has several limitations that we acknowledge in this section. 

\textbf{Study design.}
One of the most significant limitations of our research was the lack of standardization of testing. Unlike Opdahl and Sindre~\cite{opdahlExperimentalComparisonAttack2009}, where students completed assignments in a testing facility, our study consisted of students completing assignments at home with a month to complete the tasks. As such, we cannot exclude external factors from having an effect, and we could not measure data related to actual efficiency in the MEM. However, given that both populations of students were given the same conditions (training, access to resources, and time), we believe that our study design is sufficient to examine the possible effects of technical background on ADT acceptability. Additionally, this is in line with other threat model evaluation studies, such as~\cite{lallieEmpiricalEvaluationEffectiveness2017,broccia_assessing_2024}. 

As an established practice in this type of study (see Section~\ref{ssec:training}, we provided training on the method to our participants. It might be the case that the training eclipsed any innate differences between the groups. However, if this is the case, it would suggest that relatively short training is a viable means to ensure that ADTs are accessible to stakeholders with varying technical backgrounds.

Attack trees are amenable to represent physical, cyber-physical, and purely cyber scenarios~\cite{shevchenko2018threat}. The first attack tree outlined by Schneier in~\cite[Fig. 1]{schneierAttackTrees1999} represents a physical attack to open up a safe, while an attack tree from Mauw and Oostdijk captures a free lunch scenario~\cite[Fig. 1]{mauwFoundationsAttackTrees2006}.  We aim to evaluate the acceptability of ADTs outside of a domain-specific context (cyber) and our ADTs were constructed in such a way that domain knowledge is not necessary to understand them. As mentioned previously, it is recommended in the TM literature to be considerate of the used terminology to improve conveyance~\cite{ingalsbe2008threat}. However, in practice, some modeled attacks can be highly complex and require advanced security expertise. We welcome future studies that will measure the effect of the technical terms used in ADT models on the acceptability of the method for users with varying technical backgrounds.

\textbf{Participants.}
Our sample size of 102 participants in total is quite substantial and consistent with the sample sizes of similar studies evaluating threat models, which have 87~\cite{karpatiInvestigatingSecurityThreats2015}, 63~\cite{lallieEmpiricalEvaluationEffectiveness2017,opdahlExperimentalComparisonAttack2009}, 49~\cite{broccia2025evaluating}, 42~\cite{kattaComparingTwoTechniques2010}, 28~\cite{labunetsExperimentalComparisonTwo2013}, and 25~\cite{broccia_assessing_2024} participants. Still, our sample might be biased, as the participants come from the same university and the majority of them have the same country of origin. 

Another limitation of the sample is that students may not be representative of industry practitioners as a whole. Using students as study participants for threat model evaluation is standard practice with such studies~\cite{lallieEmpiricalEvaluationEffectiveness2017,labunetsExperimentalComparisonTwo2013,opdahlExperimentalComparisonAttack2009,karpatiInvestigatingSecurityThreats2015,kattaComparingTwoTechniques2010,scandariato2015descriptive}. A study by Karpati~\etal\ consisting of interviews with industry practitioners was able to confirm the results found in a previous study using student participants~\cite{karpatiComparingAttackTrees2014}, which lends itself to the idea that generally student participants can speak to the acceptability of threat models. These results were reinforced by, for example, Naiakshina~\etal~\cite{naiakshinaConductingSecurityDeveloper2020}, Salman~\etal~\cite{salmanAreStudentsRepresentatives2015}, Svahnberg~\etal~\cite{svahnbergUsingStudentsSubjects2008} and Yakdan~\etal~\cite{yakdanHelpingJohnnyAnalyze2016} who found that within the cyber security and software engineering contexts, treatment effects on computer science students hold for professionals. Based on these results, we believe that our sample of students is reflective of practitioners. 

It might be that our participants self-selected for cybersecurity-related studies, and thus, they might be more geared toward cybersecurity than the general population. This would make them more representative of a cybersecurity practitioner (who is also geared towards security) than the general population. Threat modelers will likely receive hands-on experience and training on security-related topics, and some of them might be interested in security, but not all participants in threat modeling are necessarily geared towards security~\cite{verreydt2024threat,shostack2008experiences}. Future studies should aim to examine this link with personal preferences. 

A component of our study (the large study) was graded. This might have biased the students's answers, especially regarding their perceptions, if they wanted to please the graders. We tried to mitigate this by repeatedly informing the participants that perception questions were not evaluated as a component of their grades. Additionally, our core interest is in finding differences between the two groups. If one group perceived ADTs substantially differently than the other group, we would likely still see the effect in the data. We note that some participants did report low perceptions of ADTs, and both groups did this at relatively similar rates.  

 Finally, as participation in our study was voluntary, it is possible that our students self-selected, and only students who had a high level of understanding of ADTs elected to participate in the study. 
 This is confirmed by the grade difference between the participants and non-participants as shown in Table~\ref{tab:grades}. However, we can see that the final grades between the two treatment groups are equivalent. This implies that stronger students were self-selecting in similar proportions in both cohorts, and thus there was no difference between the two groups. 
  We welcome future studies with more diverse population samples, preferably from industry practitioners that will independently examine the effect of technical (computer science) background on attack tree acceptability, especially for participants without technical background.

\section{Conclusions}\label{sec:conclusions}

ADTs are a valuable threat modeling method, recognized for its accessibility~\cite{shevchenko2018threat,reversinglabs2024attacktrees}.
We investigated whether ADTs are acceptable for users with a very limited technical background using MEM~\cite{moodyMethodEvaluationModel2003}. Overall, we find sufficient evidence to support that ADTs are equally highly acceptable for users with a very limited technical background and users with a substantial technical background. Moreover, attack trees designed by these two types of users show similar patterns in terms of the size of the trees, types of attacks modeled, and quality of the trees. We conclude that ADTs are suitable as a threat modeling method for diverse groups of stakeholders.
Further studies should look into measuring the exact effects of the technical terms on attack tree acceptability, making such models more accessible to practitioners without a technical background, and assessing different training regimens.

\subsection*{Acknowledgements}
We thank Kate Labunets, the anonymous reviewers, and our shepherd for their helpful feedback on this paper. 

This research has been partially supported by the Dutch Research Council (NWO) under the project ``Cyber Security by Integrated Design (\mbox{C-SIDe})'' (NWA.1215.18.008).


\section{Ethical Considerations}
\label{sec:ethics}

Several important ethical considerations are relevant to this research. We now outline how we considered them during the study design and execution. The Science Ethics Review Board at Leiden University reviewed and approved our study. 

Our study involved human participants, and, moreover, these participants were students taking a course taught by the authors of this paper. This introduces ethical concerns due to the dual role of the authors being both in the research team and responsible for the education of the students in the course. We have done our utmost to ensure that the students were not pressured to participate in this study and that they did not perceive being pressured or nudged to participate. Below we discuss the multiple safeguards in this regard that we introduced. 

Students were informed of the study objectives and design and then were asked to fill in and sign an informed consent form. In this form, they could choose to provide consent for their responses to the assignment to be included in the study and for the data they submit to be used for research purposes in an anonymized format. The consent forms were collected blindly for the teachers. 

We made it clear to the students that the assignment was a mandatory, graded course component, but participation in the study was entirely optional and would not affect their grades. Students were informed that teaching assistants would grade their submissions according to defined grading rubrics, and teaching assistants had no knowledge of who had elected to participate in the study. The grading rubric did not account for study participation in any way; thus, the grade was not influenced by (non-)participation. Finally, students were told that they could withdraw their consent at any time. We informed students that we would not collect responses until one month after final grades were submitted (and were no longer able to be modified). For any students who were still concerned, we offered the protocol of initially providing consent to participate, and withdrawing said consent after final grades were submitted. Withdrawing consent required filling out an online form, which we provided with the intention of making it as easy and straightforward as possible for students who no longer wished for their responses to be included.

We further provided resources when presenting the research to students, in the participation consent form, and in the introduction to the assignment that students could reach out to if they were concerned about any negative effects resulting from the study. These resources included the contact information for the relevant Ethical Review Board, the university ombudsman's office, and a student counselor. To our knowledge, no students reached out to these resources with questions or concerns about the study.

The assignments were submitted via the university's learning management software (LMS), which is a standard and accepted practice for course assignments. For the students who opted to participate in the study, once the data processing started, students were assigned a ``participant number'' which was stored in a password-protected reference list on the first author's university-issue computer. The participant number was used to anonymize the data for analysis. All other study data was pulled directly off of the LMS into a spreadsheet for further processing. The data collected and analyzed for the study did not contain any personal information.

Participants were not provided compensation for their participation in the study. As the assignment was a mandatory course component, it would have been inappropriate to compensate students for completing it. We designed our study following the Menlo Report's guidelines for ethical research~\cite{kenneallyMenloReportEthical2012} and we strived to carefully balance the benefits of the study against potential harms. The assignment itself is useful for students as it helps them learn about important concepts within cybersecurity and develops their analysis skills. We also believe that our students benefited from the study because they experienced the scientific process in the computer science domain. Moreover, the findings from this study allow us to further improve our research-based teaching, which will benefit future generations of students. It is important for the community that teachers can confidently teach attack trees to students without a substantial computer science background. Our personal experience told us that attack trees are accessible to such audiences, but only via doing a properly designed study can we be confident about this.  

We believe that the potential harm to our students, on the other hand, is limited, because we actively emphasized that non-participation does not entail any consequences for the course and we placed multiple safeguards to protect the students. Participation in the study did not entail any extra effort for the students (because they would still be doing the work as a course assignment). 

Our Ethics Review Board agreed with this risk-benefit analysis and approved our study (ref. 2022-016).

\section{Open Science}
\label{sec:open-science}

The full set of anonymized, qualitative, perception data is shared alongside this work in our supplementary data material~\cite{zenodo-dataset}. This includes all of the values used to calculate the results presented in this paper, as well as additional elements of data that were ultimately excluded.  The data are shared in a \texttt{.csv} format. To enable verification, we provide the code we used to analyze the data and generate the results presented in this paper. This code is provided as a Jupyter notebook.

Further, we provide the dataset of ADTs generated by participants. All ADTs are provided as \texttt{.png} files, with trees without structural errors provided as \texttt{.xml} files in the ADTool schema. 
Finally, we also share the slides used in the training of the study alongside a summary of the training and indicative time amounts spent on each part of the training. All these materials are available in~\cite{zenodo-dataset}.

\begin{small}
\bibliography{bibliography}{}
\bibliographystyle{myplain}
\end{small}

\appendix

\footnotesize

\section{Small Study}\label{sec:small-study-q}


\noindent\textbf{ADT 1.}
The ADT for the following questions was created by Buldas~\etal\. It can be found on page four labeled as Figure~1~\cite{buldasAttributeEvaluationAttack2020}.

\begin{itemize}
    \setlength{\itemindent}{\qsIndent}
\item[\surveyq{SS-Q2}]  How many leaf nodes are in this ADT?
\item[\surveyq{SS-Q3}]  How many root nodes are in this ADT?
\item[\surveyq{SS-Q4}]  How many different attack vectors are represented by this ADT?
\item[\surveyq{SS-Q5}]  The attack tree is easy to understand
\item[\surveyq{SS-Q6}]  I prefer this attack tree to a written description of this attack
\end{itemize}

\noindent\textbf{ADT 2.}
The ADT for the following questions is shown in Figure~\ref{img:ss-adt2}.

\begin{itemize}
    \setlength{\itemindent}{\qsIndent}
    \item[\surveyq{SS-Q7}] How many attack leaf nodes are in this ADT?
    \item[\surveyq{SS-Q8}] How many different attack vectors are represented by this ADT?
    \item[\surveyq{SS-Q9}] How many attack vectors do not have a defense?
    \item[\surveyq{SS-Q10}] The attack tree is easy to understand
    \item[\surveyq{SS-Q11}] I prefer this attack tree to a written description of this attack
\end{itemize}


\noindent\textbf{ADT 3.}
The ADT for the following questions was created by Mauw and Oostdijk~\cite{mauwRFIDCommunicationBlock}.
\begin{itemize}
    \setlength{\itemindent}{\qsIndent}
    \item[\surveyq{SS-Q12}] How many attack vectors do not have a defense?
    \item[\surveyq{SS-Q13}] How many different attack vectors are represented by this ADT?
    \item[\surveyq{SS-Q14}] How many levels of abstraction are present in this ADT?
    \item[\surveyq{SS-Q15}] The attack tree is easy to understand
    \item[\surveyq{SS-Q16}] I prefer this attack tree to a written description of this attack
\end{itemize}

\noindent\textbf{ADT 4.}
The ADT for the following questions was created by Kordy~\etal\ can be found on page 58 of that work labeled Figure~1~\cite{kordyAttackdefenseTrees2014}.

\begin{itemize}
    \setlength{\itemindent}{\qsIndent}
    \item[\surveyq{SS-Q17}] Is the overall goal kept? Why or why not?
    \item[\surveyq{SS-Q18}] How many levels of abstraction are present in this ADT?
    \item[\surveyq{SS-Q19}] The attack tree is easy to understand
    \item[\surveyq{SS-Q20}] I prefer this attack tree to a written description of this attack
\end{itemize}

\footnotesize

\section{Large Study}\label{sec:large-study-q}

\subsubsection*{ADT 1: Assembling ADTs}
The following attack \textbf{leaf} nodes are provided. The overall goal of this scenario (and thus the root node of the tree) is \textbf{Rob bank}. Assemble an attack-defense tree using these leaf nodes. Do not add any additional leaf nodes. You may add any intermediary nodes you wish.

\textbf{Attack leaf nodes:} 
Hire Outright; Promise part of the stolen money; Threaten insiders; Buy tools; Steal tools; Gain Access; Walk through front door; Locate start of tunnel; Find direction to tunnel.

\textbf{Defense leaf nodes:} 
Personnel Risk Management; Check employee financial situation.


\noindent\textbf{Likert Questions.}
\begin{itemize}
  \setlength{\itemindent}{\qIndent}
  \item[\surveyq{LS-ADT1-L1}] I find the structure of attack tree easy to understand
  \item[\surveyq{LS-ADT1-L2}] Given all the nodes of an attack tree, it is easy for me to assemble the tree
  \item[\surveyq{LS-ADT1-L3}] Given only the leaf nodes of an attack tree, it is easy for me to assemble the tree.
  \item[\surveyq{LS-ADT1-L4}] I would rather define my own intermediary nodes
  \item[\surveyq{LS-ADT1-L5}] The process of assembling the attack tree helped me better understand the attack scenario.
\end{itemize}

\noindent\textbf{Short Response Questions.}
\begin{itemize}
  \setlength{\itemindent}{\qIndent}
  \item[\surveyq{LS-ADT1-W1}] What did you find most difficult about this task? Why?
  \item[\surveyq{LS-ADT1-W2}] How did you go about solving this task? What was your methodology?
\end{itemize}

\subsubsection*{ADT 2: Building ADTs}
The following text scenario is provided for you. Please create a complete attack defense tree \textbf{of this scenario}. \textbf{Do not add extra information that is not in the scenario}. Try to encapsulate the entire scenario with an attack-defense tree (don't leave any aspect of the attack scenario out).

\emph{Scenario:} 
The goal is to open a safe. To open the safe, an attacker can pick the lock,
learn the combination, cut open the safe, or install the safe improperly so
that he can easily open it later. Some models of safes are such that they cannot be picked, so if this model is used, then an attacker is unable to pick the lock. There are also auditing services to check if safes and other security technology is installed correctly. To learn the combination, the attacker
either has to find the combination written down or get the combination
from the safe owner. If the password is such that the safe owner can remember it, then the safe owner would not need to write it down.


\noindent\textbf{Likert Questions.}
\begin{itemize}
  \setlength{\itemindent}{\qIndent}
  \item[\surveyq{LS-ADT\revised{2}-L1}] I prefer reading attack trees to text descriptions of attacks.
  \item[\surveyq{LS-ADT\revised{2}-L2}] The process of building the attack tree helped me better understand the attack scenario.
\end{itemize}

\noindent\textbf{Short Response Questions.}
\begin{itemize}
  \setlength{\itemindent}{\qIndent}
  \item[\surveyq{LS-ADT\revised{2}-W1}] What did you find most difficult about this task? Why?
  \item[\surveyq{LS-ADT\revised{2}-W2}] How did you go about building the ADT?\@ What was your methodology?
  \item[\surveyq{LS-ADT\revised{2}-W3}] What was the first node you added to your tree?
\end{itemize}

\subsubsection*{ADT 3: Creating ADTs}
Construct an attack defense tree of a scenario of your choice. Your tree should be complete (covers all reasonable attack scenarios) and reasonably large.


\noindent\textbf{Likert Questions.}
\begin{itemize}
  \setlength{\itemindent}{\qIndent}
  \item[\surveyq{LS-ADT\revised{3}-L1}] The process of creating the attack tree helped me better understand the attack scenario I selected
  \item[\surveyq{LS-ADT\revised{3}-L2}] I feel I could have achieved the same understanding by writing a text description of the attack.
  \item[\surveyq{LS-ADT\revised{3}-L3}] The ADT I created would help me communicate my threat scenario.
\end{itemize}

\noindent\textbf{Short Response Questions.}
\begin{itemize}
  \setlength{\itemindent}{\qIndent}
  \item[\surveyq{LS-ADT\revised{3}-W1}] What did you find easy about using ADTs?
  \item[\surveyq{LS-ADT\revised{3}-W2}] What did you find difficult about using ADT?\@
  \item[\surveyq{LS-ADT\revised{3}-W3}] Do you think ADTs have a place in the cybersecurity industry? If so, where? If not, why not?
  \item[\surveyq{LS-ADT\revised{3}-W4}] What aspects, if any, do you think are missing from ADTs?
  \item[\surveyq{LS-ADT\revised{3}-W5}] Do you hope to encounter ADTs in the future?
\end{itemize}

\end{document}